\documentclass[11pt,a4paper]{article}
\usepackage[paperwidth=26cm]{geometry}
\usepackage[T1]{fontenc}
\usepackage[latin1]{inputenc}
\usepackage{graphicx}
\usepackage{amsmath}
\usepackage{amssymb}

\newcommand{\bl}{\boldsymbol}

\newcommand{\lef}{\left(\,}
\newcommand{\rig}{\,\right)}
\newcommand{\ph}{\phantom}

\newcommand{\eq}{\,=\,}
\newcommand{\ma}{\,+\,}
\newcommand{\me}{\,-\,}

\begin{document}

\title{\textbf{Killing Spinors and Related Symmetries in Six Dimensions}}

\author{\textbf{Carlos Batista}\\
\small{Departamento de F\'{\i}sica}\\
\small{Universidade Federal de Pernambuco}\\
\small{50670-901 Recife-PE, Brazil}\\
\small{carlosbatistas@df.ufpe.br}}
\date{\today}




\maketitle

\begin{abstract}
Benefiting from the index spinorial formalism, the Killing spinor equation is integrated in six-dimensional spacetimes. The integrability conditions for the existence of a Killing spinor are worked out and the Killing spinors are classified in two algebraic types, in the first type the scalar curvature of the spacetime must be negative, while in the second type the spacetime must be an Einstein manifold. In addition, the equations that define Killing-Yano (KY) and closed conformal Killing-Yano (CCKY) tensors are expressed in the index notation and, as consequence,  all non-vanishing KY and CCKY tensors that can be generated from a Killing spinor are made explicit.
\\
\begin{center}
\end{center}
\end{abstract}

\section{Introduction}

It is unanimous that a great source of difficulty in general relativity, supergravity and string theories arises from the non-linear equations that plague these theories. Therefore, any method that could circumvent the problems of integrating a non-linear equation deserve ample attention in these contexts. Probably, the most fruitful way of obtaining exact solutions for the equations of motion for these theories comes from the use of symmetries. Indeed, continuous symmetries lead to conserved charges which, in turn, can generate simpler equations of motion and enable the separability of partial differential equations. An important example of this fact is given by the use of Killing vectors and Killing-Yano (KY) tensors to integrate field equations in curved backgrounds. Killing tensors and KY tensors are generally referred to as hidden symmetries, as they generate symmetry transformations in the phase space, as a whole, in contrast to Killing vectors, which generate symmetries of the space \cite{Santillan}. It turns out that Kerr-NUT-($A$)$dS$ spacetime, in arbitrary dimension, admits a closed conformal Killing-Yano (CCKY) tensor of rank two \cite{Frolov_KY}, from which one can generate just the right number of KY tensors necessary in order to fully integrate the geodesic equation \cite{Kubiz-Geodesic}, as well as Klein-Gordon \cite{Frol-KG} and Dirac equations \cite{Oota-Dirac} in such background. Killing-Yano tensors are also of special relevance in supersymmetric theories, since they can generate extra supersymmetries, as illustrates the problem of a spinning particle in a curved spacetime \cite{Gibbons_sky,Cariglia}. Moreover, KY tensors constitute true quantum symmetries, since they generate differential operators that commute with the Dirac operator \cite{KYoperator}.

In this context, spacetimes admitting Killing spinors are of particular interest, since by means of the bilinears constructed from a single Killing spinor it is possible to generate a whole tower of KY tensors. Therefore, it is of great importance to find explicit expressions for metrics admitting Killing spinors. With these motivations in mind, the main goal of this article is to use the spinorial index formalism to classify and obtain six-dimensional Lorentzian metrics of spacetimes possessing Killing spinors. Although it is definitely possible to attain several interesting results about spinors using the usual abstract notation and without fixing the dimension, it is a matter of fact that many practical results regarding manipulations of spinors depend strongly on the dimension. Since the index spinorial notation takes fully into account the dimensional specificities, this approach generally constitutes a powerful tool.
Indeed, in order to set the index spinorial notation one needs to analyse the fundamental representation of the group $Sin(\mathbb{R}^{p,q})$, whose features heavily depend on the dimension. Moreover, the use of the latter approach avoids the manipulation of the rather clumsy Fierz identities. The major example of the power of the spinorial index notation is provided by the considerable amount of results obtained by R. Penrose in four-dimensional general relativity by means of the two-component spinor formalism \cite{Penrose-Spinors-in-GR}.

Killing spinors are intimately related to the notion of holonomy group, since they can be defined as the spinors whose projective lines are invariant under the action of the holonomy group of a particular spinorial connection. Therefore, a natural question that can be posed is: which Riemannian holonomy groups are compatible with the existence of a Killing spinor? The answer has first been worked out for the case of covariantly constant spinors \cite{ParallelSpinors-Wang,ParallelSpinors-Baum}, also called parallel spinors, namely spinors that are in the kernel of the action of the holonomy group. In such a case, it turns out that the Riemannian holonomy group must be special. Differently, in the case of non-parallel Killing spinors, the Riemannian holonomy group is not necessarily special, since neither the Killing spinor nor its projective line are invariant by a parallel transport using the spinorial extension of the Levi-Civita connection. Nevertheless, it turns out that, in some cases, a non-parallel Killing spinor on a manifold $M$ can be identified with a parallel spinor on the cone over $M$ \cite{Bar} and, therefore, the cone over $M$ has special holonomy \cite{KillingSpinorsHolonomy}. This theme is of great interest in the high energy community, inasmuch as the holonomy groups and $G$-structures play an important role in supergravity and string theories \cite{Papadopoulos:2011cz}. Moreover, spaces possessing Killing spinors are of primary relevance for the latter theories, since compactifications in such spaces preserve part of the supersymmetry \cite{DuffPope-sugra}.



Besides the Holonomy approach, several other routes have been adopted in order to integrate the Killing spinor equation. Specially in supergravity
theory and string theory, in which cases the assumed connection differ from the usual Levi-Civita connection, and its spinorial extension, by terms that depend on the matter fields of these theories. In particular, these supercovariant derivatives are, generally, endowed with torsion.  In these theories, the Killing spinors are not related to the Riemannian holonomy group, but rather to the holonomy group of the new connection, which can be trickier to analyse \cite{Cacciatori_SUGRA_4D}. With no ambition of being exhaustive, in this paragraph we mention some attempts of integrating the Killing spinor equation and classifying its solutions in such context. Ref. \cite{Tod:1983pm} has made use of the spinorial index formalism in order to find four-dimensional spacetimes of minimal supergravity. In Ref. \cite{Caldarelli4D_Lorentz}, the Killing spinor equation of minimal gauged supergravity has been integrated in Lorentzian spaces of dimension four  using Fierz identities in order to generate several algebraic and differential identities involving bilinears of the Killing spinor. An analogous procedure in four-dimensional spaces of split signature has been put forward in \cite{Klemm4D_Neutral}. Similar approaches, using the bilinears built from the Killing spinors along with the Fierz relations, have also been used to find solutions of minimal supergravity in five dimensions \cite{Sugra_5D}, whereas the six-dimensional case has been considered in \cite{Gutowski:2003rg}, see also \cite{SUGRA_6D}. Concerning eleven-dimensional spacetimes, Ref. \cite{Gillard_11D} search for solutions of the Killing spinor equation by representing spinors in terms of differential forms. An elegant approach using group-theoretical tools to integrate the Killing spinor equation in symmetric manifolds have been adopted in Ref. \cite{Alberca}.

The outline of the article is the following. Sec. \ref{Sec.KillingSpinorIntro} sets some conventions adopted in the rest of the work and provides a basic review about the spinorial formalism and the Killing spinor equation. In Sec.  \ref{Sec.Spinors6D}, it is introduced the six-dimensional index spinorial formalism, which plays a central role in this work. Particularly, it is shown how differential forms and the Riemann curvature are represented in such notation. In the sequel, Sec. \ref{Sec.SymmetryEquations} shows how the equations satisfied by Killing-Yano and closed conformal Killing-Yano tensors are written in the index spinorial approach. In addition, the Killing spinor equation is also transcribed to the latter formalism. Then, in Sec. \ref{Sec.GeneratingKY}, it is constructed all possible KY and CCKY tensors that are descendant from the existence of a Killing spinor. The integrability conditions that must be satisfied by the curvature  in order for a six-dimensional manifold to admit a Killing spinor are presented in Sec. \ref{SecIntegrab.Cond.}. In the same section, a brief discussion about restrictions on the holonomy group is also addressed. The main content of this article shows up in Sec. \ref{Sec.Integration}, in which the Killing spinor equation is explicitly integrated in some particular cases. In such section, the Killing spinors in spaces of Lorentzian signature are classified in two algebraic types. Moreover, we pointed out some general prerequisites that must be satisfied by a spacetime in order for it to possess a Killing spinor of a particular type. Finally, the conclusions and prospects are presented in Sec. \ref{Sec.Conclusion}. Some supplementary material is displayed in appendices \ref{AppendixLorentz} and \ref{AppendixConnection}.

\section{Killing Spinor Equation}\label{Sec.KillingSpinorIntro}

In this section we shall set the conventions adopted throughout the article. Let $(M,\bl{g})$ be an $n$-dimensional Riemannian manifold with the metric $\bl{g}$ being of arbitrary signature. In this work we are mainly interested in local results, so that most claims are valid only in a local patch of $M$. Moreover, we shall consider solely the even-dimensional case. In what follows, $\mu$, $\nu$, $\rho$, $\cdots$ stands for coordinate indices of a local coordinate system $\{x^\mu\}$ whose associated frame of vector fields will be denoted by $\{\bl{E}_\mu= \partial_\mu\}$, these indices run from 1 to $n$. On the other hand, $a$, $b$, $c$,  $\cdots$, which also range from 1 to $n$, refers to the indices that label the vector fields $\{\bl{e}_a\}$ of a local frame such that the mutual inner products are constant,
$$  \bl{g}(\bl{e}_a,\bl{e}_b) \,\equiv\,  g_{ab}\;, \quad \textrm{where} \quad  \partial_\mu\,g_{ab} \eq 0\,.  $$
It will be assumed that $(M,\bl{g})$ admits a spin structure \cite{BennBook}, namely it has a well-defined spinorial bundle. The spinors shall be denoted by Greek letters in bold face, $\{\bl{\psi}, \,\bl{\eta},\,\cdots\}$. The Clifford action of a vector field $\bl{V}$ on a spinor field $\bl{\psi}$ shall be written as
$\bl{V}\cdot \bl{\psi}$. Since the inner products of the frame $\{\bl{e}_a\}$ are constant, it is always possible to introduce a local spinorial basis $\{\bl{\xi}_\alpha \}$, with $\alpha,\,\beta,\,\varsigma,\,\cdots$ ranging from 1 to $2^{n/2}$, such that the representation of the Clifford operators $\bl{e}_a$ are constant, namely
$$  \bl{e}_a \cdot \bl{\xi}_\alpha \eq (\gamma_a)^\beta_{\ph{\beta}\alpha}\, \bl{\xi}_\beta\,,   $$
where the matrices $\gamma_a$ are constant, $\partial_\mu (\gamma_a)^\beta_{\ph{\beta}\alpha} = 0$. In such a basis, the spinorial representation of the coordinate frame $\{\bl{E}_\mu\}$ will be denoted by the matrices $\Gamma_\mu$,
$$   \bl{E}_\mu \cdot \bl{\xi}_\alpha \eq (\Gamma_\mu)^\beta_{\ph{\beta}\alpha}\, \bl{\xi}_\beta \,. $$
Here, indices enclosed by round brackets are supposed to be symmetrized, while indices enclosed by square brackets are skew-symmetrized. Thus, for example,
\begin{equation}\label{CliffordAlgebra}
 \Gamma_{(\mu}\,\Gamma_{\nu)}   \eq  \frac{1}{2}\lef    \Gamma_{\mu}\,\Gamma_{\nu} \ma \Gamma_{\nu}\,\Gamma_{\mu} \rig \eq  g_{\mu\nu} \,,
\end{equation}
which stems from the very definition of Clifford algebra. Analogously,
$$ \Gamma_{\mu\nu} \,\equiv\, \Gamma_{[\mu}\, \Gamma_{\nu]}\eq \frac{1}{2}\lef    \Gamma_{\mu}\,\Gamma_{\nu} \me \Gamma_{\nu}\,\Gamma_{\mu} \rig \,,$$
where we have adopted the usual definition $\Gamma_{\mu_1\cdots \mu_p} \equiv \Gamma_{[\mu_1}\cdots \Gamma_{\mu_p ]}$.

Since we are considering the case in which the dimension $n$ is even, we can introduce the chirality matrix
$$ \Upsilon\,\propto\, \gamma_1 \gamma_2 \cdots \gamma_n\,, $$
where the proportionality constant is chosen in a way that $\Upsilon^2 \eq 1$. This matrix splits the spinor bundle in two subbundles of dimension $2^{\frac{n-2}{2}}$, according to the eigenvalue of the spinor with respect its action. Spinors with eigenvalue $1$ are called positive chirality Weyl spinors, while those with eigenvalue $-1$ are called Weyl spinors of negative chirality. Besides the chirality matrix, there are some other matrices that play an important role in the spinorial formalism. For instance, invertible matrices $A$, $B$ and $C$ obeying the relations
$$  A\,\gamma_a\,A^{-1} \eq -\,\gamma_a^\dag \quad, \quad    B\,\gamma_a\,B^{-1} \eq \gamma_a^{\,*} \quad, \quad  C\,\gamma_a\,C^{-1} \eq -\,\gamma_a^t\,, $$
where $\gamma_a^\dag$, $\gamma_a^*$ and $\gamma_a^t$ stands for the hermitian conjugate, the complex conjugate and the transpose matrix of $\gamma_a$, respectively. In particular, note that we can set $C = BA$. The matrix $B$ enables the definition of a basis-independent notion of complex conjugation of a spinor. Indeed, if $\bl{\psi}$ is a spinor, its charge conjugate is defined by
$$      \overline{\bl{\psi}} \eq   B^{-1}\,\bl{\psi}^* \;, $$
namely, its components are $\overline{\psi}^{\,\alpha} \eq (B^{-1})^\alpha_{\ph{\alpha}\beta} \,\, (\psi^{\beta})^*$. The main feature of this definition is that $\overline{\bl{\psi}}$ and $\bl{\psi}$ transform in the same way under the connected part (to the identity) of the orthogonal group that preserves the metric $g_{ab}$. Further, by means of the matrix $C$, one can define an inner product between spinors that is also invariant under the action of the latter group:
\begin{equation}\label{inner1}
  \langle \bl{\psi}, \bl{\phi} \rangle \eq \bl{\psi}^t\,C\, \bl{\phi} \eq \psi^\alpha\,C_{\alpha\beta}\,\phi^\beta  \,.
\end{equation}
Using this product one can, in addition, define the following other products that, likewise, are invariant under the action of the connected part of the orthogonal group:
\begin{equation}\label{inner2}
\langle \bl{\psi},\Upsilon\, \bl{\phi} \rangle \quad, \quad   \langle \overline{\bl{\psi}}, \bl{\phi} \rangle
\quad, \quad   \langle \overline{\bl{\psi}}, \Upsilon\, \bl{\phi} \rangle   \,.
\end{equation}
Generally, all these inner products are different and independent. In particular, the inner product $ \langle \bl{\psi},\Upsilon\, \bl{\phi} \rangle$ is invariant under the action of the group $Pin(\mathbb{R}^n)$, the universal covering of the group $O(n)$.   Note that since $\langle \overline{\bl{\psi}}, \overline{\bl{\phi}} \rangle =  \pm \langle \bl{\psi}, \bl{\phi}  \rangle^{*} $, it follows that $\langle \overline{\bl{\psi}}, \overline{\bl{\phi}} \rangle$ is not independent from  $\langle \bl{\psi}, \bl{\phi} \rangle$. Actually, no further inner products can be generated in a trivial way.

Let $\nabla$ denote the Levi-Civita connection of the tangent bundle. Then, this connection can be uniquely extended to the spinorial bundle if we assume that its extension is compatible with the Clifford action as well as with the natural inner products on the spinorial bundle \cite{BennBook,Batista-PureSubspace}. We shall denote this extension by the same symbol. More precisely, the following Leibniz rules are assumed to hold:
\begin{equation}\label{LeibnizClifford}
  \nabla_a\,\left(\bl{V} \cdot \bl{\psi} \right)\eq   \left(\nabla_a\,\bl{V}  \right)\cdot \bl{\psi} \ma \bl{V} \cdot \nabla_a\,\bl{\psi}
\quad \textrm{ and }\quad     \nabla_a\,\langle\bl{\psi},\bl{\phi}\rangle \eq   \langle\nabla_a\,\bl{\psi},\bl{\phi}\rangle \ma    \langle\bl{\psi},\nabla_a\,\bl{\phi}\rangle  \,,
\end{equation}
where $\bl{V}$ is an arbitrary vector field while $\bl{\psi}$ and $\bl{\phi}$ are arbitrary spinor fields. If the covariant derivatives of the frame vectors $\{\bl{e}_a\}$ are given by
\begin{equation}\label{TangentConnection}
  \nabla_a \, \bl{e}_b \eq \omega_{ab}^{\ph{ab}c}\,\bl{e}_c
\end{equation}
then one can show that, in order for Eq. (\ref{LeibnizClifford}) to hold, the spinorial connection must have the following action in the spinorial frame:
\begin{equation}\label{CovDSpinors}
  \nabla_a \, \bl{\xi}_\alpha \eq (\Omega_{a})^\beta_{\ph{\beta}\alpha}\,\bl{\xi}_\beta\quad , \quad \quad \textrm{where}\quad  \Omega_{a} \eq  -\,\frac{1}{4}\,\omega_a^{\ph{a}bc}\, \gamma_b \,\gamma_c \,.
\end{equation}
Thus, if $\bl{\psi}\eq \psi^\alpha \,  \bl{\xi}_\alpha$ is a general spinor then
$$ \nabla_a \, \bl{\psi} \eq \lef \partial_a  \psi^\alpha \ma (\Omega_{a})^\alpha_{\ph{\alpha}\beta}\,\psi^\beta \rig \bl{\xi}_\alpha \,. $$
Using indices, as done henceforth, the latter equation is written as
$$  \nabla_a \,  \psi^\alpha \eq \partial_a \,\psi^\alpha \ma (\Omega_{a})^\alpha_{\ph{\alpha}\beta}\,\psi^\beta \,.     $$
Additionally, the covariant derivative of an object with lower spinorial index, namely a section in the dual spinor bundle, is given by
$$ \nabla_a   \zeta_\alpha \eq \partial_a \,\zeta_\alpha \me (\Omega_{a})^\beta_{\ph{\beta}\alpha}\,\zeta_\beta \,.  $$
The derivative of objects with several spinorial indices follows trivially from the latter equations. For instance,
\begin{equation}
  \nabla_a N^\alpha_{\ph{\alpha}\beta} \eq  \partial_a N^\alpha_{\ph{\alpha}\beta} \ma (\Omega_{a})^\alpha_{\ph{\alpha}\varsigma}\, N^\varsigma_{\ph{\alpha}\beta} \me N^\alpha_{\ph{\alpha}\varsigma} \, (\Omega_{a})^\varsigma_{\ph{\alpha}\beta}  \quad \Rightarrow\quad
 \nabla_a N\eq \partial_a N \ma [\Omega_a, N]\,,
\end{equation}
where the spinorial indices have been omitted in the latter identity and $[\Omega_a, N]$ stands for the commutator of the ``matrices'' $\Omega_a$ and $N$. In particular, since $\partial_a \gamma_b = 0$, it follows that the derivative of the Clifford operators $\gamma_a$ are given by:
$$ \nabla_a \gamma_b \eq  [\Omega_a, \gamma_b] \eq \omega_{ab}^{\ph{ab}c}\,\gamma_c \,,$$
which resembles Eq. (\ref{TangentConnection}), as it should. On the other hand, when computing the covariant derivative of $\Gamma_\mu$, one must account for the covariant derivative of both the vectorial index $\mu$ as well as the omitted spinorial indices. Pleasantly, it turns out that these contributions cancel each other, so that
$$ \nabla_\mu \, \Gamma_\nu \eq 0 \,.  $$
Further, from the definition of the operators $A$, $B$ and $C$, the following relations can be easily obtained
$$  A\,\Omega_a \eq -\, \Omega_a^{\,\dag}\,A \quad, \quad  B\,\Omega_a \eq  \Omega_a^{\,*}\,B \quad, \quad  C\,\Omega_a \eq -\, \Omega_a^{\,t}\,C  \,. $$
These properties are of relevance for proving that the defined extension of the Levi-Civita connection satisfy the Leibniz rule with respect to the inner products defined in Eqs. (\ref{inner1}) and (\ref{inner2}). Finally, before proceeding, let us point out two important identities encompassing the Riemann curvature of the manifold. First, it is possible to relate the curvature of the spinorial connection to the Riemann tensor as follows \cite{BennBook}
\begin{equation}\label{SpinorCurvature}
  \left( \nabla_\mu\,\nabla_\nu   \me   \nabla_\nu\,\nabla_\mu\right) \bl{\psi}   \eq \frac{1}{4}\, R_{\mu\nu}^{\ph{\mu\nu}\rho\sigma}\,
\Gamma_{\rho\sigma}\, \bl{\psi}\,.
\end{equation}
Further, if $\mathcal{D} = \Gamma^\mu\,\nabla_\mu$ is the Dirac operator, then, manipulating the latter relation, it follows that
\begin{equation}\label{DiracSquared}
  \mathcal{D}^2\,\bl{\psi} \eq  \nabla^\mu\,\nabla_\mu \bl{\psi} \me  \frac{1}{4}\,R\,\bl{\psi}  \,,
\end{equation}
where $R$ stands for the Ricci scalar.

A non-zero spinor field $\bl{\psi}$ is said to be a Killing spinor whenever there exists a constant $\alpha$ such that the following equation holds:
\begin{equation}\label{KillingSpinor}
 \nabla_\mu \, \bl{\psi} \eq \alpha \, \Gamma_\mu\, \bl{\psi} \,.
\end{equation}
We shall call the constant $\alpha$ the eigenvalue of the Killing spinor, since it is, essentially, the eigenvalue under the action of the Dirac operator. Indeed, Eq. (\ref{KillingSpinor}) trivially implies $\frac{1}{n}\mathcal{D}\,\bl{\psi} = \alpha\,\bl{\psi}$. Now, taking the covariant derivative of the above equation and using Eq. (\ref{SpinorCurvature}), it is straightforward to arrive at the following integrability condition:
\begin{equation}\label{IntegrabilityCond}
 R_{\mu\nu}^{\ph{\mu\nu}\rho\sigma}\,\Gamma_{\rho\sigma}\, \bl{\psi} \eq -  8\, \alpha^2 \Gamma_{\mu\nu}  \bl{\psi}\,.
\end{equation}
The latter relation turns out to be equivalent to the following integrability conditions:
\begin{equation}\label{IntegrabCond2}
  R \eq -\, 4\, \alpha^2\, n(n-1) \quad,\quad  \lef R_{\mu\nu} - \frac{R}{n}\,g_{\mu\nu} \rig \Gamma^\nu\, \bl{\psi} \eq  0 \quad,\quad
  C_{\mu\nu\rho\sigma}\,\Gamma^{\rho\sigma}\, \bl{\psi} \eq 0 \,.
\end{equation}
Where $R_{\mu\nu}$ is the Ricci tensor while $C_{\mu\nu\rho\sigma}$ denotes the conformal curvature, the so-called Weyl tensor. In particular, the first of these relations imply that the constant $\alpha$ must be either real o purely imaginary, depending on the sign of the scalar curvature. Further, in a space whose Ricci scalar vanishes, the Killing spinor must be covariantly constant. Moreover, applying $\Gamma_\sigma$ on the left of the second identity in Eq. (\ref{IntegrabCond2}) and using the fact that in the Euclidean signature the matrices $\Gamma_\mu$ admit an hermitian representation, one can check that in, this signature, the manifold is an Einstein manifold \cite{Nieuwenhuizen}. Nonetheless, in non-Euclidean signatures, it is perfectly possible for a non-Einstein manifold to admit a Killing spinor.

From the relation $\Upsilon\,\Gamma_\mu = - \Gamma_\mu\, \Upsilon$, valid in even dimensions, and from the fact that $\Upsilon$ is covariantly constant it follows that, given a Killing spinor $\bl{\psi}$, one can construct another Killing spinor whose eigenvalue have opposite sign. More precisely, if $\bl{\psi}$ obeys Eq. (\ref{KillingSpinor}) then
$$ \widetilde{\bl{\psi}} \,\equiv\, \Upsilon\,\bl{\psi} \quad \Rightarrow \quad
 \nabla_\mu \, \widetilde{\bl{\psi}} \eq -\, \alpha \, \Gamma_\mu\, \widetilde{\bl{\psi}} \,. $$
Therefore, in even dimensions, Killing spinors with non-zero eigenvalue ($\alpha\neq 0$), come in pairs with their eigenvalues differing by the sign. In particular, this implies that a Killing spinor with non-zero eigenvalue cannot be chiral. Now, let us see that if $\bl{\psi}$ is a Killing spinor with eigenvalue $\alpha$ then its conjugate, $\overline{\bl{\psi}}$, is a Killing spinor with eigenvalue $\alpha^*$. In order to attain this result, one must realize that $\nabla_\mu \bl{\psi}^* \neq (\nabla_\mu \bl{\psi})^*$, which stems from the fact that the spinorial connection $\Omega_a$ is not real. Indeed, one can easily check that
$$  (\nabla_\mu \bl{\psi})^* \eq  \nabla_\mu (\bl{\psi}^*)  \ma (\Omega_\mu^{\,*} \me \Omega_\mu )\,\bl{\psi}^*  \,. $$
Thus, assuming that $\bl{\psi}$ obeys Eq. (\ref{KillingSpinor}), it follows that
\begin{align*}
  \nabla_\mu\,\overline{\bl{\psi}} &\eq   \nabla_\mu\,( B^{-1}\, \bl{\psi}^*) \eq (\Omega_\mu\,B^{-1} \me  B^{-1}\, \Omega_\mu ) \bl{\psi}^* \ma B^{-1}\, \nabla_\mu (\bl{\psi}^*)  \\
  & \eq      B^{-1}\,\left[ \,(B\,\Omega_\mu\,B^{-1} \me  \Omega_\mu ) \bl{\psi}^* \ma \nabla_\mu (\bl{\psi}^*) \right] \eq
  B^{-1}\,\left[ \,(\Omega_\mu^{\,*} \me  \Omega_\mu ) \bl{\psi}^* \ma \nabla_\mu (\bl{\psi}^*) \right]\\
   & \eq    B^{-1}\,(\nabla_\mu \bl{\psi})^* \eq  B^{-1}\,\lef \alpha \, \Gamma_\mu\, \bl{\psi} \rig^* \eq \alpha^* \, \Gamma_\mu\, \overline{\bl{\psi}}\,,
\end{align*}
which is the desired result.

It is simple matter to prove that every Killing spinor obeys the twistor equation,
\begin{equation}\label{twistorEq}
  \nabla_\mu \, \bl{\psi} \eq \frac{1}{n}\,\Gamma_\mu\,\mathcal{D}\,\bl{\psi}  \,.
\end{equation}
However, the converse is not true in general. For instance, a twistor can be chiral, while a Killing spinor with non-zero eigenvalue cannot. Nonetheless, in the special case of an Einstein manifold, one can always use a twistor to generate a Killing spinor \cite{Nieuwenhuizen}. More precisely, if $\bl{\chi}$ is a twistor in an Einstein manifold then the spinors
$$ \bl{\psi}_{\pm} \, \equiv\,  \bl{\chi} \,\pm\,  \sqrt{\frac{4(n-1)}{-\,n\,R}} \,\,\mathcal{D}\,\bl{\chi}$$
are Killing spinors with eigenvalues
\begin{equation}\label{AlphaR}
  \alpha \eq \pm \, \sqrt{\frac{-\,R}{4\,n\,(n-1)}} \,.
\end{equation}
However, it is worth stressing that, generally,  for manifolds that do not obey the Einstein condition, the existence of a twistor does not imply the existence of a Killing spinor.

One of the main practical utilities of the Killing spinors is that they can be used to generate symmetry tensors which, eventually, lead to conservation laws as well as to the integrability of field equations using the manifold as a background. Indeed, if $\bl{\psi}$ is a Killing spinor then each of the tensors
\begin{equation}\label{KilligYanos}
\langle \bl{\psi},\Gamma_{\mu_1\mu_2\cdots \mu_p} \bl{\psi} \rangle \quad, \quad
 \langle  \bl{\psi} ,\Gamma_{\mu_1\mu_2\cdots \mu_p}\,\Upsilon\, \bl{\psi} \rangle \quad, \quad
 \langle \overline{ \bl{\psi}},\Gamma_{\mu_1\mu_2\cdots \mu_p} \bl{\psi} \rangle \quad \textrm{and} \quad
 \langle \overline{\bl{\psi}},\Gamma_{\mu_1\mu_2\cdots \mu_p} \,\Upsilon\,\bl{\psi} \rangle
\end{equation}
is either a Killing-Yano (KY) tensor or a closed conformal Killing-Yano (CCKY) tensor.
For instance, let us take the derivative of the first of these skew-symmetric tensors:
\begin{align}
 \nabla_\nu\, \langle \bl{\psi},\Gamma_{\mu_1\cdots \mu_p} \bl{\psi} \rangle &\eq
\langle \alpha\, \Gamma_\nu\, \bl{\psi},\Gamma_{\mu_1\cdots \mu_p} \bl{\psi} \rangle \ma
\langle  \bl{\psi},\Gamma_{\mu_1\cdots \mu_p} \alpha\, \Gamma_\nu\, \bl{\psi} \rangle \nonumber\\
&\eq \alpha \, \bl{\psi}^t\,\lef  \Gamma_\nu^t \,C\, \Gamma_{\mu_1\cdots \mu_p} \ma C \,\Gamma_{\mu_1\cdots \mu_p}\,\Gamma_\nu  \rig\, \bl{\psi} \nonumber\\
&\eq \alpha \, \bl{\psi}^t\,C\, \lef -\, \Gamma_\nu\, \Gamma_{\mu_1\cdots \mu_p} \ma \Gamma_{\mu_1\cdots \mu_p}\,\Gamma_\nu \rig\, \bl{\psi} \label{KS-KY}
\end{align}
Then, by means of the Clifford algebra, one can prove that
$$ -\, \Gamma_\nu\, \Gamma_{\mu_1 \mu_2\cdots \mu_p} \ma \Gamma_{\mu_1 \mu_2 \cdots \mu_p}\,\Gamma_\nu \eq
\left\{                                                     \begin{array}{ll}
                                                                                                -\,2\,\Gamma_{\nu\mu_1 \mu_2\cdots\mu_p} \quad \textrm{if $p$ is odd\,,}  \\
\quad\\
                                                                                                  -\,2\,p\,g_{\nu[\mu_1}\Gamma_{\mu_2\cdots\mu_p]} \quad \textrm{if $p$ is even\,.}
                                                                                                \end{array}
                                                                                              \right.
 $$
Therefore, defining
$$ Y_{\mu_1 \mu_2 \cdots\mu_p} \,\equiv\, \langle \bl{\psi},\Gamma_{\mu_1 \mu_2 \cdots \mu_p} \bl{\psi} \rangle \,, $$
it follows from Eq. (\ref{KS-KY}) that if $p$ is odd then
$$ \nabla_\nu\, Y_{\mu_1\mu_2\cdots\mu_p} \ma \nabla_{\mu_1}\, Y_{\nu\mu_2\cdots\mu_p} \eq  0\,, $$
namely the tensor $\bl{Y}$ is a Killing-Yano tensor. On the other hand, if $p$ is even the following equation holds
$$  \nabla_\nu\, Y_{\mu_1\mu_2\cdots\mu_p} \eq 2\, g_{\nu[\mu_1}\,h_{\mu_2\cdots\mu_p]} \;, \quad \textrm{where}\quad
h_{\mu_1\cdots\mu_{p-1}} \,\equiv\,  -\,p\, \langle \bl{\psi},\Gamma_{\mu_1\cdots \mu_{p-1}} \bl{\psi} \rangle \,,  $$
which means that $\bl{Y}$ is a closed conformal Killing-Yano tensor or, equivalently, its Hodge dual is a Killing-Yano tensor. The proof that the other tensors in (\ref{KilligYanos}) are either a KY or a CCKY tensor goes in a completely analogous fashion. However, the tensors in (\ref{KilligYanos}) are not all independent from each other. Indeed, apart from a multiplicative constant, the Hodge dual of
$\langle \bl{\psi},\Gamma_{\mu_1\cdots \mu_p} \bl{\psi} \rangle$ is equal to
$\langle \bl{\psi},\Gamma_{\nu_1\cdots \nu_{n-p}} \Upsilon \bl{\psi} \rangle$, so that they both generate the same conservation laws. Moreover, some of these tensors might be identically zero. It seems that one of the first works to explicitly construct symmetry tensors using Killing spinors was Ref. \cite{ParallelSpinors-Wang}, in which covariantly constant spinors were used to generate the covariantly constant differential forms that define Euclidean manifolds of special holonomy. Conformal Killing-Yano tensors generated by twistors have been considered in Ref. \cite{Semmelmann:2002fra} and their generalizations for connections with totally skew-symmetric torsion have been investigated in Ref.  \cite{Houri-Torsion}. The use of these symmetry tensors in the context of supergravity have also been contemplated in the literature \cite{Cariglia}.


\section{Spinors in Six Dimensions}\label{Sec.Spinors6D}

Now its time to move on to the specific case of interest in this article, namely six-dimensional manifolds. Although one can obtain several relevant results about spinors without specifying the dimension, as illustrated in the preceding section, it turns out that many features of spinors are highly dependent on the dimension. Therefore, it is always fruitful and illuminating to stick to a particular dimension and explore its particularities in full detail. Indeed, this was the path taken by R. Penrose  when he  introduced the two-component spinors in four dimensions \cite{Penrose-Spinors-in-GR}. Unarguably, this two-component formalism played an important role on the understanding of several results in four-dimensional general relativity. Given this motivation, the aim of this work is to take advantage of the spinorial index notation valid specifically in six dimensions in order to exploit the subject of Killing spinors. A detailed introduction to the spinorial index formalism  in six dimensions is available in Ref. \cite{Spin6D}. Previous accounts of such formalism have also appeared in the literature in the context of conformal field theories \cite{Weinberg-Conf6D,Mason6D}, see also \cite{Batista_Conf6D}. Moreover, this approach have been used to investigate the Kerr theorem in six dimensions \cite{Kerr6D}. The index spinorial formalism in five dimensions have also been worked out in the literature \cite{Spinor5D}, see also \cite{DeSmet}.

The key to find how spinors should be represented in six dimensions is to note that $Spin(\mathbb{R}^6)$, the double covering of $SO(6)$, is isomorphic to $SU(4)$. Therefore, a chiral spinor must transform in a 4-dimensional representation of $SU(4)$. There are two such representations:  the representation that associates to each element  $\bl{U}\in SU(4)$ the usual $4\times 4$ unitary matrix $U^A_{\ph{A}B}$ of unit determinant; and the representation that associates to $\bl{U}\in SU(4)$  a matrix $\tilde{U}_{A}^{\ph{A}B}$ that is the transposed  inverse and of $U^A_{\ph{A}B}$. Where the indices $A,\,B,\,C,\,\cdots$ range from 1 to 4. All other 4-dimensional representations are related to one of these two by similarity transformations. Thus, the objects $\psi^A$ and $\psi_A$ that transform under the action of $\bl{U}\in SU(4)$ as
\begin{equation}\label{SU(4)-fund.}
  \psi^{A}\,\stackrel{\bl{U}}{\longrightarrow} \,U^A_{\phantom{A}B} \,\psi^{B}\;\;\;\;\; ,\, \;\;\;\;\; \psi_{A}\,\stackrel{\bl{U}}{\longrightarrow}\, \tilde{U}_{A}^{\ph{A}B}\, \psi_{B}
\end{equation}
are the desired chiral spinors. More precisely, we shall say that $\psi^A$ is a spinor of positive chirality while $\psi_A$ has negative chirality. Therefore, a general spinor, also called a Dirac spinor, is given by a pair $\bl{\psi}=(\psi^A, \psi_A)$. It is worth stressing that, generally, the components $\psi^A$ and $\psi_A$ are totally independent from each other, summing 8 complex degrees of freedom  for the Dirac spinor $\bl{\psi}$. Note that the scalar $\psi^A \psi_A$ is invariant under the action of $SU(4)$ and, thus, invariant under $Spin(\mathbb{R}^6)$.

Now, let us define the symbols $\varepsilon_{ABCD}$ and $\varepsilon^{ABCD}$ to be such that
$$ \varepsilon_{ABCD} \eq \varepsilon_{[ABCD]} \;\;,\;\; \varepsilon_{1234} \eq 1 \quad\textrm{ and }\quad
\varepsilon^{ABCD} \eq \varepsilon^{[ABCD]} \;\;,\;\; \varepsilon^{1234} \eq 1 \,. $$
Then, due to the fact that the matrices $U^A_{\ph{A}B}$ and $\tilde{U}_{A}^{\ph{A}B}$ have unit determinant, it follows that these totally anti-symmetric symbols are invariant under the actions of $SU(4)$. Therefore, these objects might be related to the metric of the Euclidean space, as the metric is invariant under the action of $SO(6)$. Due to this relation, one concludes that the spinorial representation of a vector $V^a$ in six dimensions is given by an object with two spinorial indices that are skew-symmetric, $V^{AB} \eq V^{[AB]}$, so that the following relation holds
$$  g_{ab}\,V^a \,Z^b \eq  \frac{1}{2}\,\varepsilon_{ABCD} \, V^{AB}\,Z^{CD}  \,,$$
where the factor of $1/2$ was chosen for convenience. Note that the objects  $V^{AB}=V^{[AB]}$ have 6 degrees of freedom, just as the vectors $V^a$. In the same fashion as one can use the metric $g_{ab}$ and its inverse $g^{ab}$ to lower or raise vectorial indices, a pair of skew-symmetric spinorial indices can also be lowered or raised by means of the symbols $\varepsilon_{ABCD}$ and $\varepsilon^{ABCD}$:
\begin{equation}\label{MetricSpinor}
 V_a\eq  g_{ab}\,V^b   \quad \Leftrightarrow \quad  V_{AB}\,=\,\frac{1}{2} \, \varepsilon_{ABCD}\,V^{CD} \quad\quad \textrm{and }
\quad\quad V^a\eq  g^{ab}\,V_b   \quad \Leftrightarrow \quad  V^{AB}\,=\,\frac{1}{2} \, \varepsilon^{ABCD}\,V_{CD} \,.
\end{equation}
It is worth stressing that, differently from the 4-dimensional case, in six dimensions there is no natural way to raise or lower a \emph{single} spinorial index.

Just as $SO(6)$ vectors are represented in the spinorial formalism by  $V^{AB}=V^{[AB]}$, other tensors also transform according to specific spinorial representations. For instance, let us obtain the spinorial representation of a bivector, namely a skew-symmetric tensor of rank two, $B_{ab}=B_{[ab]}$. From what we have just seen, it follows that the bivectors in six dimensions are represented in the spinorial formalism by objects $\mathfrak{B}_{AB\,CD}$ such that
$$ \mathfrak{B}_{AB\,CD} \eq  \mathfrak{B}_{[AB]\,[CD]} \quad \textrm{ and } \quad \mathfrak{B}_{AB\,CD}  \eq -\, \mathfrak{B}_{CD\,AB} \,.    $$
However, one can check that these objects are in one-to-one correspondence with the objects of the form $B^A_{\ph{A}B}$ that have vanishing trace, $B^A_{\ph{A}A}=0$. Indeed, the ``isomorphism'' map is given by $B^A_{\ph{A}B} = \varepsilon^{ADCE}\,\mathfrak{B}_{DC\,EB}$. Therefore, we can say that, in the six-dimensional spinorial formalism, bivectors are trace-less objects with one index up and one index down. Table \ref{Table spinors equivalent} compiles the spinorial representation for some tensors of interest. An analogous table is available in Ref. \cite{Spin6D}, to which the reader is directed to in order to grasp its details.

\begin{table}[!htbp]
\begin{center}
\begin{tabular}{c c c}
\hline \hline
  $SO(6)$ \textsc{Rep.} \quad\quad\quad\;& \textsc{Spinorial Rep.} & \textsc{Algebraic Symmetries}   \\ \hline
  $V^a$ & $V^{AB}$ & $V^{AB}=V^{[AB]}$ \\
  $S_{ab}\,=\,S_{(ab)}\,,\,S^a_{\ph{a}a}=0$ &$ S^{AB}_{\phantom{AB}CD}$ &\quad\quad  $S^{AB}_{\phantom{AB}CD}=S^{[AB]}_{\phantom{AB}[CD]}, \, S^{AB}_{\phantom{AB}CB}=0$ \\
  $B_{ab}\,=\,B_{[ab]}$ & $B^A_{\phantom{A}B}$ & $B^A_{\phantom{A}A}=0$ \\
  $T_{abc}\,=\,T_{[abc]}$  & $(T^{AB},T_{AB})$& $T^{AB}=T^{(AB)}, T_{AB}=T_{(AB)}$  \\
  $C_{abcd} = C_{[cd][ab]}\,,\,C^a_{\;\,\,bad}=0$ & $C^{AB}_{\phantom{AB}CD}$ & \quad\quad$C^{AB}_{\phantom{AB}CD}=C^{(AB)}_{\phantom{AB}(CD)},  C^{AB}_{\phantom{AB}CB} = 0$ \\
  \hline\hline
\end{tabular}
\caption{\small The first column gives the tensorial representation according to $SO(6)$. For instance, $S_{ab}$ denotes a traceless symmetric tensor of rank two, while the object $C_{abcd}$ is any tensor with the same algebraic symmetries of a Weyl tensor. The second column gives the spinorial representation for each tensor of the first column. The third column gives the algebraic symmetries that must be satisfied by the spinorial indices.} \label{Table spinors equivalent}
\end{center}
\end{table}
\normalsize

In what follows, we shall use the basis $\{\bl{\chi}_p\}$ for expressing the spinors of positive chirality, where the components of the spinors of this basis are
$$ \bl{\chi}_1 \,\leftrightarrow\, \delta_1^{\,A} \quad, \quad  \bl{\chi}_2 \,\leftrightarrow\,\delta_2^{\,A} \quad, \quad
 \bl{\chi}_3 \,\leftrightarrow\, \delta_3^{\,A} \quad, \quad  \bl{\chi}_4 \,\leftrightarrow\, \delta_4^{\,A} \,, $$
while the basis adopted for spinors of negative chirality will be the dual basis $\{\bl{\zeta}^p\}$, whose components are
$$ \bl{\zeta}^1 \,\leftrightarrow\, \delta^1_{\,A}  \quad, \quad  \bl{\zeta}^2 \,\leftrightarrow\, \delta^2_{\,A} \quad, \quad
\bl{\zeta}^3 \,\leftrightarrow\, \delta^3_{\,A} \quad, \quad \bl{\zeta}^4 \,\leftrightarrow\, \delta^4_{\,A} \,. $$
By means of these bases we can build the following independent vector fields that form a frame for the tangent bundle
$$ e_1^{\,AB} \eq \chi_1^{\,[A}\chi_2^{\,B]} \quad,\quad e_2^{\,AB} \eq \chi_1^{\,[A}\chi_3^{\,B]} \quad,\quad  e_3^{\,AB} \eq \chi_1^{\,[A}\chi_4^{\,B]} $$
\begin{equation}\label{NullFrame1}
  e_4^{\,AB} \eq \chi_3^{\,[A}\chi_4^{\,B]} \quad,\quad e_5^{\,AB} \eq \chi_4^{\,[A}\chi_2^{\,B]} \quad,\quad  e_6^{\,AB} \eq \chi_2^{\,[A}\chi_3^{\,B]} \,.
\end{equation}
Lowering the skew-symmetric pairs of spinorial indices, we find that
$$ e_{1\,AB} \eq \zeta^3_{\,[A}\zeta^4_{\,B]} \quad,\quad e_{2\,AB} \eq \zeta^4_{\,[A}\zeta^2_{\,B]} \quad,\quad  e_{3\,AB} \eq \zeta^2_{\,[A}\zeta^3_{\,B]} $$
\begin{equation}\label{NullFrame2}
  e_{4\,AB} \eq \zeta^1_{\,[A}\zeta^2_{\,B]} \quad,\quad e_{5\,AB} \eq \zeta^1_{\,[A}\zeta^3_{\,B]} \quad,\quad  e_{6\,AB} \eq \zeta^1_{\,[A}\zeta^4_{\,B]} \,.
\end{equation}
The inner products of the vector fields of this basis are:
\begin{equation}\label{NullFrameInnerP}
  \bl{g}(\bl{e}_i,\bl{e}_j) \eq 0 \quad ,\quad \bl{g}(\bl{e}_i,\bl{e}_{j+3}) \eq \frac{1}{2}\, \delta_{i\,j} \quad  ,
\quad \bl{g}(\bl{e}_{i+3},\bl{e}_{j+3}) \eq 0\,,
\end{equation}
with the indices $i$ and $j$ running from 1  to 3. In particular, all vector fields in this basis are null, so that we shall refer to it as a null frame. The reality conditions of this frame are related to the signature of the metric \cite{Trautman}. For instance, in the Lorentzian case the reality condition is given by \cite{Bat-Book}:
\begin{equation}\label{Reality_Vectors}
  \bl{e}_1^{\,*} \eq  \bl{e}_1 \quad, \quad \bl{e}_4^{\,*} \eq  \bl{e}_4 \quad, \quad  \bl{e}_2^{\,*} \eq  \bl{e}_5 \quad, \quad
\bl{e}_3^{\,*} \eq  \bl{e}_6 \,,
\end{equation}
which can be easily verified by means of constructing a null frame out of a real Lorentz frame whose metric is $\textrm{diag}(-1,1,1,1,1,1 )$. In order for the reality condition (\ref{Reality_Vectors}) to hold, one can assume that the charge conjugation of the spinors in the basis are given by:
\begin{align}
 \overline{\bl{\chi}}_1 \eq & \bl{\chi}_2 \quad, \quad \overline{\bl{\chi}}_2 \eq -\, \bl{\chi}_1 \quad, \quad  \overline{\bl{\chi}}_3 \eq -\,\bl{\chi}_4 \quad, \quad \overline{\bl{\chi}}_4 \eq \bl{\chi}_3 \label{Conjugation}\\
 \overline{\bl{\zeta}}^1 \eq & \bl{\zeta}^2 \quad, \quad \overline{\bl{\zeta}}^2 \eq -\,\bl{\zeta}^1 \quad, \quad   \overline{\bl{\zeta}}^3 \eq -\,\bl{\zeta}^4 \quad, \quad \overline{\bl{\zeta}}^3 \eq \bl{\zeta}^4 \,.   \nonumber
\end{align}
Adopting these conjugation rules is tantamount to assume Lorentz signature, which can be checked by mens of using (\ref{NullFrame1}) along with (\ref{Conjugation}) and then comparing with (\ref{Reality_Vectors}).


In what follows, the covariant derivative of a Dirac spinor  $\bl{\psi}=(\psi^A,\psi_A)$ will be written as
\begin{equation}\label{Cov Deriv Spinor6D}
   \nabla_{AB}\,\psi^C \eq \partial_{AB} \,\psi^C \ma (\Omega_{AB})^C_{\ph{C}D}\,\psi^D \quad \;\textrm{and} \quad\;
\nabla_{AB}\,\psi_C \eq \partial_{AB} \,\psi_C \me \psi_D\,(\Omega_{AB})^D_{\ph{D}C} \,,
\end{equation}
where $(\Omega_{AB})^C_{\ph{C}D}$ is the spinorial connection and $\partial_{AB}$ stands for the partial derivative. For instance, $\partial_{34}$ denotes the partial derivative along the vector field $\bl{e}_4$, in accordance with Eq. (\ref{NullFrame1}). For more details about this spinorial connection, see appendix \ref{AppendixConnection}.


Since the Riemann tensor is skew-symmetric in its two pairs of indices, $R_{abcd}=R_{[ab][cd]}$, it follows, from the bivector representation shown in Table \ref{Table spinors equivalent}, that the Riemann tensor is given in the spinorial formalism by $\mathfrak{R}^{A\ph{A}C}_{\ph{A}B\ph{C}D}$, where $\mathfrak{R}^{A\ph{A}C}_{\ph{A}A\ph{C}D}=0$ and $\mathfrak{R}^{A\ph{A}C}_{\ph{A}B\ph{C}C}=0$. Thus, the analogue of Eq. (\ref{SpinorCurvature}) in this index notation is \cite{Kerr6D,Batista_Conf6D}:
\begin{equation}\label{SpinorCurvature1}
  \varepsilon^{GABC}\,(\nabla_{AB}\,\nabla_{CD}\me \nabla_{CD}\,\nabla_{AB})\,\psi^E  \eq 4\,\mathfrak{R}^{G\ph{D}E}_{\ph{A}D\ph{E}F}\,\psi^F  \,.
\end{equation}
The torsionless property of the connection imply that the action of the curvature operator on the scalar $\chi^E\lambda_{E}$ must vanish, from which one concludes that
\begin{equation}\label{SpinorCurvature2}
  \varepsilon^{GABC}\,(\nabla_{AB}\,\nabla_{CD}\me \nabla_{CD}\,\nabla_{AB})\,\psi_E  \eq-\, 4\,\mathfrak{R}^{G\ph{D}F}_{\ph{A}D\ph{F}E}\,\psi_F \,.
\end{equation}
Just as the Riemann tensor can be decomposed in its irreducible parts with respect to the local action of the orthogonal group, the same can be done for its spinorial analogue. More precisely, this decomposition is given by
\begin{equation}\label{SpinCurv-Riemann}
   4\,\mathfrak{R}^{A\ph{C}B}_{\ph{A}C\ph{B}D} \eq \Psi^{AB}_{\phantom{AB}CD} \ma \Phi^{AB}_{\phantom{AB}CD}
 \me \frac{1}{60}\,R \left(\, \delta^A_C\,\delta^B_D  \me 4\, \delta^A_D\,\delta^B_C\,\right) \,,
\end{equation}
where $\Psi^{AB}_{\phantom{AB}CD}$ represents the Weyl tensor, $\Phi^{AB}_{\phantom{AB}CD}$ stands for the traceless part of the Ricci tensor and $R$ is the Ricci scalar. In accordance with Table \ref{Table spinors equivalent}, the algebraic symmetries of the latter objects are the following:
\begin{equation}\label{Algeb.Curvature}
  \Psi^{AB}_{\phantom{AB}CD}=\Psi^{(AB)}_{\phantom{AB}(CD)}\quad,\quad \Phi^{AB}_{\phantom{AB}CD}=\Phi^{[AB]}_{\phantom{AB}[CD]} \quad ,\quad \Psi^{AB}_{\phantom{AB}CB} \eq 0 \eq  \Phi^{AB}_{\phantom{AB}CB} \,.
\end{equation}

Now that we have set the conventions and introduced the spinorial index formalism in six dimensions, it is time to present the main results of this work. To the best of authors' knowledge, the results of the following sections are original and have not appeared elsewhere.






\section{Symmetry Equations in Six Dimensions}\label{Sec.SymmetryEquations}

The goal of this section is to obtain how the equations obeyed by Killing-Yano (KY) tensors, closed conformal Killing-Yano (CCKY) tensors and Killing spinors are transcribed to the spinorial formalism using the index notation presented in the previous section.

The cases of a Killing vector field $\bl{Y}$ and a closed conformal Killing vector $\bl{H}$ are quite simple. These fields obey the following equations respectively
$$  \nabla_a\,Y_b \ma \nabla_b\,Y_a \eq 0 \quad \textrm{and} \quad \nabla_a\,H_b \eq 2\,h\,g_{ab} \,, $$
where $h$ is some scalar function. Therefore, if $Y_{AB}$ and $H_{AB}$ are the spinorial representations of these fields, it is immediate to conclude that the spinorial equations satisfied by these fields are
\begin{equation}\label{KVector}
  \nabla_{AB}\,Y_{CD} \ma   \nabla_{CD}\,Y_{AB} \eq 0   \quad \textrm{and}  \quad   \nabla_{AB}\,H_{CD} \eq h\,\varepsilon_{ABCD}  \,,
\end{equation}
where the equivalence $g_{ab}\sim \frac{1}{2}\varepsilon_{ABCD}$, that comes from Eq. (\ref{MetricSpinor}), has been taken into account. An equivalent way to write the equation satisfied by a Killing vector field is $\nabla_a Y_b = B_{ab}$, where $B_{ab}=B_{[ab]}$ is some bivector. By lack of any other option, due to the necessary arrangement of indices, the spinorial analogue of the latter equation must be given by
$$  \nabla_{AB}\,Y^{CD} \,\propto\, \delta^{[C}_{\ph{\;}[A} \, B^{D]}_{\ph{D}B]} \,,$$
where the traceless object $B^{D}_{\ph{D}B}$ is the spinorial representation of the bivector $B_{ab}$, see Table \ref{Table spinors equivalent}. The proportionality multiplicative factor missing in the previous relation can be chosen to be 1 by absorbing it into the bivector $B^{D}_{\ph{D}B}$.

Now, let us obtain how the equation of a CCKY tensor of rank two is written in the spinorial formalism. If $\bl{H}$ is a closed conformal Killing-Yano bivector then there exists some 1-form $h_c$ such that the following equation holds
\begin{equation}\label{CCKYBiv1}
\nabla_{a}\,H_{bc} \eq 2\,g_{a[b}\,h_{c]}\,.
\end{equation}
In particular, the exterior derivative of $\bl{H}$ vanishes. If $H^A_{\ph{A}B}$ and $h_{AB}$ are the spinorial representations of $H_{ab}$ and $h_a$ respectively, then the most general way to arrange the spinorial indices to assure that the derivative of $H^A_{\ph{A}B}$ is linear in $h_{AB}$  is
$$   \nabla_{AB} \,H^C_{\ph{C}D} \eq a\, \varepsilon_{ABDE}\,h^{CE} \ma  b\, \delta^C_{\;[A}\,h_{B]D} \ma c\, \delta^C_{\;D}\,h_{AB}\,, $$
where $a$, $b$ and $c$ are constants. However, due to the identity
$$ \varepsilon_{ABDE}\,V^{CE} \eq  2\,  \delta^C_{\;[A}\,V_{B]D} \ma \delta^C_{\;D}\,V_{AB} \,, $$
valid for any vector $V_{AB}$, it follows,  without loss of generality, that one can set $a=0$ while redefining $b$ and $c$. Assuming this and noting that $H^C_{\ph{C}C}=0$, one concludes that $b=4c$. Then, the remaining multiplicative constant can be absorbed in $h_{AB}$. In conclusion, if $H^A_{\ph{A}B}$ is a CCKY bivector then there exists some $h_{AB}=h_{[AB]}$ such that the following relation holds:
\begin{equation}\label{CCKYBiv2}
\nabla_{AB} \,H^C_{\ph{C}D} \eq  \delta^C_{\;[A}\,h_{B]D} \ma \frac{1}{4}\, \delta^C_{\;D}\,h_{AB}\,.
\end{equation}
In particular, the latter equation implies that $\nabla_{A(B}H^A_{\ph{A}C)}=0$ and $\nabla^{A(B}H^{C)}_{\ph{A)}A}=0$ which, holding simultaneously, means that the exterior derivative of the bivector $H^A_{\ph{A}B}$ vanishes \cite{Batista_Conf6D}, as it should.

On the other hand, if $\bl{Y}$ is a bivector obeying the KY equation then there exists some 3-vector $\bl{T}$ such that
$$ \nabla_{a}\,Y_{bc} \eq T_{abc}\,. $$
Thus, the covariant derivative of $\bl{Y}$ is proportional to a 3-vector $\bl{T}$. In the spinorial formalism, a 3-vector is represented by a pair of symmetric objects with two spinorial indices, $(T^{AB},T_{AB})$, where $T^{AB}$ is related to the self-dual part of the 3-vector, while $T_{AB}$ represents the anti-self-dual part. The bivector $Y_{ab}$ is represented by the traceless object $Y^A_{\ph{A}B}$. Therefore, in order for the expression $\nabla_{AB} \,Y^C_{\ph{C}D}$ to be skew-symmetric in the pair of indices $AB$ and traceless in the pair $CD$, one must have
\begin{equation}\label{KYBivector}
  \nabla_{AB} \,Y^C_{\ph{C}D} \eq 2\,\delta^C_{\;[A}\,T_{B] D} \ma \varepsilon_{ABDE}\, T^{EC}\,.
\end{equation}
The freedom on the choice of the constant coefficients that could appear have been absorbed in the definition of $T^{AB}$ and $T_{AB}$.  As a consistence check, one can verify that the divergence of $Y^C_{\ph{C}D}$ vanishes as it should. Indeed, the condition of vanishing divergence is given by
$\nabla_{A[B} Y^A_{\ph{A}C]} = 0$ \cite{Batista_Conf6D}, which is satisfied whenever Eq. (\ref{KYBivector}) holds, since  $T^{AB}$ and $T_{AB}$ are both symmetric. At this point is worth stressing that just as the components $\psi^A$ and $\psi_A$ of a Dirac spinor $\bl{\psi}=(\psi^A,\psi_A)$ are totally independent from each other, the components $T^{AB}$ and $T_{AB}$ of a general 3-vector $\bl{T}=(T^{AB},T_{AB})$ are also independent. Indeed, $T^{AB}$ and $T_{AB}$ have 10 components each, which sums up the 20 components of a 3-vector in six dimensions.

Now, if $\bl{H}$ is a closed conformal Killing-Yano tensor of rank three, then it follows that there exists some bivector $\bl{h}$ such that
$$ \nabla_a \, H_{bcd} \eq 2\,g_{a[b}\,h_{cd]} $$
Thus, if $(H^{AB},\, H_{AB})$ is the spinorial representation of the 3-vector  $\bl{H}$, and $h^A_{\ph{A}B}$ is proportional to the spinorial representation of the bivector $\bl{h}$, then it follows, by lack of other algebraic possibilities, that
$$  \nabla_{AB}\, H^{CD} \eq a\, \delta^{(C}_{\;[A}\, h^{D)}_{\ph{D}B]} \quad \textrm{and} \quad
\nabla^{AB}\, H_{CD} \eq b\, \delta^{[A}_{\;(C}\, h^{B]}_{\ph{B}D)}  \,. $$
In order for the 3-vector $\bl{H}$ to have vanishing exterior derivative one must set $a = b$, so that $\nabla_{AB}H^{AC}= \nabla^{AC}H_{AB}$ \cite{Batista_Conf6D}. Moreover, the remaining freedom on the collective multiplicative factor can be absorbed in the definition of $h^A_{\ph{A}B}$. Thus, $(H^{AB},\, H_{AB})$ is a CCKY of rank three if, and only if, there exists some bivector $h^A_{\ph{A}B}$ such that
\begin{equation}\label{CKY3vec}
   \nabla_{AB}\, H^{CD} \eq  \delta^{(C}_{\;[A}\, h^{D)}_{\ph{D}B]} \quad \textrm{and} \quad
\nabla^{AB}\, H_{CD} \eq  \delta^{[A}_{\;(C}\, h^{B]}_{\ph{B}D)} \,.
\end{equation}
Since every KY tensor is the Hodge dual of some CCKY, and since the Hodge dual of the 3-vector $(H^{AB},\, H_{AB})$ is the 3-vector $(H^{AB},\, -H_{AB})$, we conclude that if $\bl{Y}=(Y^{AB},Y_{AB})$ is a KY 3-vector it satisfies the equation
\begin{equation}\label{KY3vec}
   \nabla_{AB}\, Y^{CD} \eq  \delta^{(C}_{\;[A}\, \textsf{B}^{D)}_{\ph{D}B]} \quad \textrm{and} \quad
\nabla^{AB}\, Y_{CD} \eq  -\,\delta^{[A}_{\;(C}\,  \textsf{B}^{B]}_{\ph{B}D)}
\end{equation}
for some bivector $\textsf{B}^{A}_{\ph{A}B}$. Apart from a multiplicative constant, such bivector is the Hodge dual of the exterior derivative of $\bl{Y}$. Note that it is pointless to present the spinorial equations satisfied by KY and CCKY tensors of higher ranks, since the latter are just the Hodge dual of one of the previous tensors and, therefore, do not lead to new conservation laws. For example, a KY tensor of rank four is the Hodge dual of a CCKY tensor of rank two, while a CCKY tensor of rank four is the Hodge dual of a rank two KY tensor.

Finally, let us obtain how the Killing spinor equation is represented in the index formalism. In order to accomplish this, one must recognize that the Clifford action of a vector field $\bl{V}$ in a spinor $\bl{\psi} = (\psi^A, \psi_A)$ is given by
\begin{equation}\label{CliffordAction6D}
  \bl{V} \cdot \bl{\psi} \eq \lef 2\,V^{AB}\,\psi_B \,,\,  -\,2\,V_{AB}\,\psi^B \rig \,,
\end{equation}
as one can check noting that in such a way the Clifford algebra of Eq. (\ref{CliffordAlgebra}) is properly satisfied. Then, if $\bl{\psi}=(\psi^A,\psi_A)$ is a Killing spinor of eigenvalue $\alpha$, namely $\nabla_a \bl{\psi}= \alpha \,\gamma_a \bl{\psi}$,  one concludes that
\begin{equation}\label{KillinSp6D1}
  \nabla_a \, \psi^C \eq 2\,\alpha\,(e_a)^{CD}\,\psi_D \quad  \textrm{ and } \quad  \nabla^a \, \psi_C \eq -\,2\,\alpha\,(e^{\,a})_{\;CD}\,\psi^D \,.
\end{equation}
Thus, since the vectorial index $a$ can be replaced by a skew-symmetric pair of spinorial indices, and since $(e_a)^\mu (e^b)_\mu \eq \delta^b_a$, the following equivalences hold:
$$ (e_a)^{AB} \,\leftrightarrow \,  (e_{_{CD}})^{AB} \eq \delta^{[A}_{\;C}  \, \delta^{B]}_{\;D}  \quad , \quad
(e^a)_{AB} \,\leftrightarrow \,  (e^{_{CD}})_{AB} \eq \delta^{[C}_{\;A}  \, \delta^{D]}_{\;B}  \,. $$
Hence, the Killing spinor equation can be written as
\begin{equation}\label{KilliingSpinor6D1}
   \nabla_{AB} \, \psi^C \eq 2\,\alpha\,\delta^C_{\;[A}\,\psi_{B]} \quad  \textrm{ and } \quad
\nabla^{AB} \, \psi_C \eq -\,2\,\alpha\,\delta^{[A}_{\;C}\,\psi^{B]} \,.
\end{equation}
An equivalent form of writing these equations can be obtained by means of raising and lowering the derivative indices in Eq. (\ref{KilliingSpinor6D1}), after which one obtain the following equations that might be  satisfied by a Killing spinor:
\begin{equation}\label{KilliingSpinor6D2}
  \nabla^{AB} \, \psi^C \eq  \alpha\, \varepsilon^{ABCD}\,\psi_D \quad  \textrm{ and } \quad
  \nabla_{AB} \, \psi_C \eq -\,\alpha\, \varepsilon_{ABCD}\,\psi^D  \,.
\end{equation}
The Table \ref{TableEquations} sums up the spinorial form of the symmetry equations that we have seen in this section.

\begin{table}[!htbp]
\begin{center}
\begin{tabular}{l l l}
\hline \hline
  \textsc{Type of Symmetry} & \;\textsc{Equations Satisfied} & \textsc{Constraints}  \\ \hline
  KY of rank 1 &  \;$ \nabla_{AB}\,Y^{CD} \eq \delta^{[C}_{\ph{\;}[A} \, B^{D]}_{\ph{D}B]} $ &  $B^{A}_{\ph{A}A}=0$ \\
  KY of rank 2 &  \;$\nabla_{AB} \,Y^C_{\ph{C}D} = 2\,\delta^C_{\;[A}\,T_{B] D} + \varepsilon_{ABDE}\, T^{EC}$  & $T^{AB}= T^{(AB)}$\;;\; $T_{AB}= T_{(AB)}$\\
  KY of rank 3 &  \;$ \nabla_{AB}\, Y^{CD} = \delta^{(C}_{\;[A}  \textsf{B}^{D)}_{\ph{D}B]}$ \;;\;  $\nabla^{AB}\, Y_{CD} = -\delta^{[A}_{\;(C} \textsf{B}^{B]}_{\ph{B}D)}$ &  $ \textsf{B}^{A}_{\ph{A}A}=0$ \\
CCKY of rank 1 &   \;$ \nabla_{AB}\,H_{CD} \eq h\,\varepsilon_{ABCD}$ & \;\; $\times$ \;\; \\
CCKY of rank 2 & \;$\nabla_{AB} \,H^C_{\ph{C}D} \eq  \delta^C_{\;[A}\,h_{B]D} \ma \frac{1}{4}\, \delta^C_{\;D}\,h_{AB}$ & $h_{AB} = h_{[AB]}$ \\
CCKY of rank 3 &  \;$ \nabla_{AB}\, H^{CD} = \delta^{(C}_{\;[A} h^{D)}_{\ph{D}B]}$ \;;\;  $\nabla^{AB}\, H_{CD} = \delta^{[A}_{\;(C} h^{B]}_{\ph{B}D)}$ &  $h^{A}_{\ph{A}A}=0$ \\
Killing Spinor &  \;$ \nabla_{AB} \, \psi^C \eq 2\,\alpha\,\delta^C_{\;[A}\,\psi_{B]} $ \;;\; $\nabla^{AB} \, \psi_C \eq -\,2\,\alpha\,\delta^{[A}_{\;C}\,\psi^{B]}$ & \;\; $\times$ \;\;\\
  \hline\hline
\end{tabular}

\caption{\small The first column presents the type of tensor, where KY stands for Killing-Yano tensor and CCKY denotes a closed conformal Killing-Yano tensor. In the second column, it is shown the spinorial equations satisfied by the objects of the first column. The third column displays the constraints that might be satisfied by the objects that appear in the right hand side of the spinorial equations. } \label{TableEquations}
\end{center}
\end{table}
\normalsize

There are some advantages of writing the preceding symmetry equations using spinorial indices. For example, the integrability conditions necessary for these equations to admit a solution are generally easier to obtain and understand in this spinorial formalism, as has proved to be the case in four dimensions. Moreover, these equations might be more easily integrated. For instance, in four-dimensional spacetimes, when one assumes that the Petrov type of the Weyl tensor is $D$ it is immediate to verify that there are two null directions that play a distinguished role and that some Newman-Penrose coefficients vanish, which facilitates the integration of Einstein's vacuum equation. Indeed, due to the latter fact, Kinnersley have been able to fully integrate Einstein's vacuum equation for type $D$ spacetimes \cite{Kinnersly-typeD}. Particularly, in this article, we will exploit the index formalism introduced above in order to integrate the Killing spinor equation.

\section{Constructing Symmetry Tensors from a Killing Spinor}\label{Sec.GeneratingKY}

An useful profit of using the index notation introduced in the previous sections is that one can easily guess wether a symmetry can be used to generate another symmetry. The spinorial indices are like small building blocks that can be combined in order to construct tensors that might carry a relevant geometrical significance. In this section we will illustrate this fact constructing symmetry tensors out of a Killing spinor, in the spirit of Eq. (\ref{KilligYanos}).

Let $\bl{\psi} = (\psi^A,\psi_A)$ be a Killing spinor with eigenvalue $\alpha$,  namely Eqs. (\ref{KilliingSpinor6D1}) and (\ref{KilliingSpinor6D2}) are satisfied. The first combination that we can form with this spinor is the scalar $\psi^A \psi_A$. Computing its derivative, we have
$$ \nabla_{AB}\, (\psi^C \, \psi_C) \eq  2\,\alpha\,\delta^C_{\;[A}\,\psi_{B]}\,\psi_C \me \alpha\, \psi^C\,\varepsilon_{ABCD}\,\psi^D \eq 0 \,. $$
Therefore, the scalar $\psi^A \psi_A$ must be constant. Next, note that one can use the Killing spinor to build the object $\psi^A\psi_B$ that resembles a bivector. However, generally, it is not traceless. But, subtracting its trace we end up with $\mathcal{Y}^A_{\ph{A}B}\eq \psi^A\psi_B - \frac{1}{4}(\psi^C  \psi_C)\delta^A_B$ which, is, indeed, a bivector. Then, taking its covariant derivative, we find
$$\nabla_{CD} \, \mathcal{Y}^A_{\ph{A}B}\eq \nabla_{CD} \lef \psi^A\psi_B - \frac{1}{4}\,\psi^E  \psi_E\,\delta^A_B \rig  \eq  \nabla_{CD} \lef \psi^A\psi_B\rig \eq
2\,\alpha\,\delta^A_{\;[C}\,\psi_{D]}\,\psi_B  \me \alpha\, \,\varepsilon_{CDBE}\,\psi^A\,\psi^E \,. $$
Comparing the latter relation with Eq. (\ref{KYBivector}), we conclude that $\mathcal{Y}^A_{\ph{A}B}$ is a Killing-Yano bivector whose exterior derivative is proportional to the 3-vector $(T^{AB},T_{AB}) = (\psi^A\psi^B,-\psi_A\psi_B)$. Finally, using the Killing spinor one can construct 3-vectors of the form
$(c_{_1} \psi^A\psi^B, c_{_2} \psi_A\psi_B)$, where $c_{_1}$ and $c_{_2}$ are constants. Computing the derivative of this 3-vector we find that
$$  \nabla_{AB}\lef \psi^C\,\psi^D \rig  \eq  c_{_1}\, \delta^{(C}_{\;[A}\,\mathcal{Y}^{D)}_{\ph{D}B]}  \quad \textrm{ and } \quad
\nabla^{AB}\lef \psi_C\,\psi_D \rig  \eq  -\,c_{_2}\, \delta^{[A}_{\;(C}\,\mathcal{Y}^{B]}_{\ph{B}D)} \,.$$
Therefore, taking $c_1=c_2 = 1$ we conclude that $(\psi^A\psi^B,\psi_A\psi_B)$ is a KY 3-vector, while its Hodge dual   $(\psi^A\psi^B,-\,\psi_A\psi_B)$, obtained by taking  $c_1=-c_2 = 1$, is a CCKY tensor of rank 3.

Note that, although we have been able to construct KY tensors of rank 2 and 3 using the Killing spinor $(\psi^A, \psi_B)$, no vector field has been built. Indeed, it is not hard to get convinced that there is no natural way to construct a vector using just one Killing spinor, since a vector has a skew-symmetric pair of spinorial indices and this cannot be designed using just the indices of one spinor along with the natural objects $\varepsilon_{ABCD}$ and $\delta^A_{\;B}$. In particular, this observation implies that, in six dimensions, the relations
$$ \langle \bl{\psi}\,,\, \Gamma_\mu\, \bl{\psi}\rangle  \eq 0  \quad \textrm{ and } \quad
\langle \bl{\psi}\,,\, \Gamma_\mu\, \Upsilon\, \bl{\psi}\rangle  \eq 0     $$
are valid for any spinor $\bl{\psi}$, which can be checked after introducing a representation for the matrices $\gamma_a$. This example illustrates very well how the index notation adopted here can be used to anticipate results that, otherwise, would not be obvious.

Although we cannot build a Killing vector field using just the Killing spinor $\bl{\psi}$, it is possible to construct vector fields if we use the conjugated spinor, $\overline{\bl{\psi}}$, along with the spinor $\bl{\psi}$ itself. Indeed, if $(\psi^A, \psi_A)$ is a Killing spinor with eigenvalue $\alpha$ and $(\overline{\psi}^A, \overline{\psi}_A)$ is its conjugate, then it is obvious that the following vector fields can be built:
$$ \mathcal{V}^{AB} \eq  \psi^{[A}\,\overline{\psi}^{B]} \quad \textrm{ and } \quad \widetilde{\mathcal{V}}_{AB} \eq  \psi_{[A}\,\overline{\psi}_{B]}   \,.$$
Despite these vector fields not being Killing vectors nor closed conformal Killing vectors, it is straightforward to check that the following combinations of them are
$$  \mathcal{K}_{\pm}^{\;AB}  \eq \mathcal{V}^{AB} \,\mp \,  \widetilde{\mathcal{V}}^{AB} \,. $$
By the same token, one can also define the following useful scalars, Bivectors and 3-vectors respectively,
$$ F_{\pm} \eq \psi^E\, \overline{\psi}_{E} \,\mp\, \overline{\psi}^{E}\, \psi_{E} \,, $$
$$ \mathcal{B}^{\,\,A}_{\pm\,\,B} \eq \lef  \psi^A\, \overline{\psi}_{B} \,\pm\, \overline{\psi}^{A}\,  \psi_B\rig
\me \frac{1}{4}\,\delta^A_{\;B}\, \lef \psi^C\, \overline{\psi}_{C} \,\pm\, \overline{\psi}^{C}\,  \psi_C \rig  \,,$$
$$ \lef \mathcal{T}_{\pm}^{AB} , \mathcal{T}^{\pm}_{AB} \rig  \eq   \lef \psi^{(A}\,\overline{\psi}^{B)}\, ,\, \pm \,\psi_{(A}\,\overline{\psi}_{B)} \rig \,. $$
In the case of a real eigenvalue $\alpha$, one can verify, after some algebra, that these objects obey the following differential equations:
\begin{align*}
\nabla_{AB}\, F_{+} &\eq  -\,4\,\alpha\, \mathcal{K}_{-\,AB} \\
\nabla_{AB}\, F_{-} &\eq  0 \\
  \nabla_{AB}\, \mathcal{K}_{+}^{\;CD} &\eq -\,4\,\alpha\, \delta^{[C}_{\ph{\;}[A} \, \mathcal{B}^{\,\,D]}_{-\,B]} \,,  \\
 \nabla_{AB}\, \mathcal{K}_{-\,CD} &\eq -\,\frac{\alpha}{2} \,F_{+}\, \varepsilon_{ABCD} \,,\\
 \nabla_{AB}\, \mathcal{B}^{\,\,C}_{+\,\,D} &\eq -\,2\,\alpha \lef  2\,\delta^C_{\;[A}\,\mathcal{T}^-_{B] D} \ma
\varepsilon_{ABDE}\, \mathcal{T}_-^{EC}  \rig  \,,  \\
\nabla_{AB}\, \mathcal{B}^{\,\,C}_{-\,\,D} &\eq -\,4\,\alpha \lef \delta^C_{\;[A}\,\mathcal{K}_{+\,B]D} \ma \frac{1}{4}\,
 \delta^C_{\;D}\,\mathcal{K}_{+\,AB} \rig \,, \\
\nabla_{AB}\, \mathcal{T}_{\pm}^{CD} &\eq 2\,\alpha\,\delta^{(C}_{\;[A}  \mathcal{B}^{\,D)}_{+\,B]} \quad ,\quad
  \nabla^{AB}\, \mathcal{T}^{\pm}_{CD} \eq \mp\,2\,\alpha\,\delta^{[A}_{\;(C} \mathcal{B}^{\,B]}_{+\,D)} \,.
\end{align*}
These equations, along with Table \ref{TableEquations}, imply that, when $\alpha$ is real, the tensors $\bl{\mathcal{K}_+}$, $\bl{\mathcal{B}_+}$ and $\bl{\mathcal{T}_+}$ are Killing-Yano tensors of rank 1, 2 and 3 respectively, whereas the tensors $\bl{\mathcal{K}_-}$, $\bl{\mathcal{B}_-}$ and $\bl{\mathcal{T}_-}$ are closed conformal Killing-Yano tensors of rank 1, 2 and 3 respectively. Moreover, the scalar $F_-$ is constant. On the other hand, if the eigenvalue $\alpha$ of the Killing spinor $\bl{\psi}$ is purely imaginary then the roles of these tensors are interchanged, namely $\bl{\mathcal{K}_-}$, $\bl{\mathcal{B}_-}$ and $\bl{\mathcal{T}_-}$ are Killing-Yano tensors while $\bl{\mathcal{K}_+}$, $\bl{\mathcal{B}_+}$ and $\bl{\mathcal{T}_+}$ are closed conformal Killing-Yano tensors. Moreover, for imaginary $\alpha$, the scalar $F_+$ is constant, whereas $F_-$ ceases to be constant. The results of this section are summed up in Table \ref{TableSymmetryTensors}.
\begin{table}[!htbp]
\begin{center}
\begin{tabular}{l l l}
\hline \hline
  \textsc{Type of Symmetry} & \;\textsc{Spinorial Form}  \\ \hline
Constant Scalar & \; $\psi^A\,\psi_A $ \\
  KY of rank 2 &  \;$ \psi^A\psi_B - \frac{1}{4}(\psi^C  \psi_C)\delta^A_B$ \\
  KY of rank 3 &  \;$(\psi^A\, \psi^B, \psi_A\, \psi_B )$\\
  CCKY of rank 3 &  \;$(\psi^A\, \psi^B, -\,\psi_A\, \psi_B )$\\
Constant Scalar & \; $\psi^A\, \overline{\psi}_{A} \,+\, \frac{\,\alpha^*}{\alpha}\, \overline{\psi}^{A}\, \psi_{A}$ \\
  KY of rank 1 &   \; $\psi^{[A}\,\overline{\psi}^{B]} \me \frac{\,\alpha^*}{2\alpha}\, \varepsilon^{ABCD} \,\psi_C\,\overline{\psi}_D $ \\
  KY of rank 2 &   \; $   \left( \psi^A\, \overline{\psi}_{B} \,+\,\frac{\,\alpha^*}{\alpha}\, \overline{\psi}^{A}\,  \psi_B \right)
\me \frac{1}{4}\,\delta^A_{\;B}\, \lef \psi^C\, \overline{\psi}_{C} \,+\,\frac{\,\alpha^*}{\alpha}\, \overline{\psi}^{C}\,  \psi_C \rig  $ \\
  KY of rank 3 &   \; $\left(\, \psi^{(A}\,\overline{\psi}^{B)}\, ,\, \frac{\,\alpha^*}{\alpha}\, \,\psi_{(A}\,\overline{\psi}_{B)} \,\right) $ \\
 CCKY of rank 1 &   \; $\psi^{[A}\,\overline{\psi}^{B]} \ma \frac{\,\alpha^*}{2\alpha}\, \varepsilon^{ABCD} \,\psi_C\,\overline{\psi}_D $ \\
 CCKY of rank 2 &   \; $   \left( \psi^A\, \overline{\psi}_{B} \,-\,\frac{\,\alpha^*}{\alpha}\, \overline{\psi}^{A}\,  \psi_B \right)
\me \frac{1}{4}\,\delta^A_{\;B}\, \lef \psi^C\, \overline{\psi}_{C} \,-\,\frac{\,\alpha^*}{\alpha}\, \overline{\psi}^{C}\,  \psi_C \rig  $ \\
 CCKY of rank 3 &   \; $\left(\, \psi^{(A}\,\overline{\psi}^{B)}\, ,\, -\,\frac{\,\alpha^*}{\alpha}\, \,\psi_{(A}\,\overline{\psi}_{B)} \,\right) $ \\
  \hline\hline
\end{tabular}

\caption{\small This table describes the symmetry tensors that can be constructed out of a Killing spinor $\bl{\psi}= (\psi^A, \psi_A)$ with eigenvalue $\alpha$. In the above table $\overline{\bl{\psi}}= (\overline{\psi}^A, \overline{\psi}_A)$  is the charge conjugate of $\bl{\psi}$. Here, KY stands for Killing-Yano tensor and CCKY denotes a closed conformal Killing-Yano tensor. Since $\overline{\bl{\psi}}$ is also a Killing spinor, it follows that the second, third and fourth objects of this table remain being symmetry tensors if we replace $\bl{\psi}$ by $\overline{\bl{\psi}}$. } \label{TableSymmetryTensors}
\end{center}
\end{table}
\normalsize


\section{Integrability Conditions for the Killing Spinor Equation}\label{SecIntegrab.Cond.}

In the present section, we shall obtain the integrability conditions for the Killing spinor equation. Suppose that $\bl{\psi}= (\psi^A, \psi_A)$ is a Killing spinor with eigenvalue $\alpha$, so that Eq. (\ref{KilliingSpinor6D1}) holds. Then, due to the identity (\ref{SpinorCurvature1}), it follows that
\begin{align*}
  4\,\mathfrak{R}^{G\ph{D}F}_{\ph{A}D\ph{F}E}\,\psi^E &\eq 2\, (\nabla^{GC}\,\nabla_{CD}\me \nabla_{CD}\,\nabla^{GC})\,\psi^F  \\
 & \eq 4\,\alpha^2\,  \delta^{[G}_{D}\,\psi^{F]} \me 6\,\alpha^2\,\delta^{F}_{D}\,\psi^G \,.
\end{align*}
Now, taking the symmetric and skew-symmetric parts of this equation in the pair of indices $GF$,  we eventually arrive at the following integrability conditions that must be satisfied by the curvature in order for the space to admit a Killing spinor:
\begin{equation}\label{IntegrabCondSpin1}
  R \eq -\, 120\,\alpha^2  \quad,\quad \;   \Phi^{AB}_{\phantom{AB}CD}\,\psi^D \eq 0      \quad,\quad \;  \Psi^{AB}_{\phantom{AB}CD}\,\psi^D \eq 0 \,.
\end{equation}
These constraints are the counterparts  of  the integrability conditions presented in Eq. (\ref{IntegrabCond2}). Likewise, working out the integrability condition associated to the negative chirality part of the Killing spinor, it follows that
\begin{equation}\label{IntegrabCondSpin2}
  \Phi^{AB}_{\phantom{AB}CD}\,\psi_B \eq 0      \quad,\quad \;  \Psi^{AB}_{\phantom{AB}CD}\,\psi_B \eq 0 \,.
\end{equation}
As explicitly demonstrated in the following section, these integrability conditions represent strong constraints on the types of manifolds that support a Killing spinor field.

It is well-known that the Lie algebra of the holonomy group associated to a connection is determined by the curvature of this connection, this is the content of the Ambrose-Singer theorem. More precisely, if $\bl{\mathcal{R}}^\alpha_{\ph{\alpha}\beta}$ is the curvature 2-form of some connection on a fiber bundle, and if and $\bl{B}$ is a simple bivector tangent to a plane in a tangent space, then the interior product of the latter into the former, $i_{\bl{B}}\bl{\mathcal{R}}^\alpha_{\ph{\alpha}\beta}$, is the generator of the holonomy transformation associated to an infinitesimal loop contained in the plane tangent to $\bl{B}$. In the case of the spinorial bundle of a six-dimensional space, the action of the Lie algebra generator associated to the parallel transport of a spinor along an infinitesimal loop in the plane tangent to the bivector $B^A_{\ph{A}B}$ can be written in the index formalism as follows:
$$  \lef \phi^A\,,\,\phi_A \rig \;\mapsto\; \lef B^{D}_{\ph{D}C}\,\mathfrak{R}^{C\ph{D}A}_{\ph{C}D\ph{F}B}\,\phi^B \,,\,
    -\,B^{D}_{\ph{D}C}\,\mathfrak{R}^{C\ph{D}B}_{\ph{C}D\ph{F}A}\,\phi_B \rig \,. $$
Therefore, using the integrability conditions obtained in the present section along with Eq. (\ref{SpinCurv-Riemann}), it follows that the action of the previous Lie algebra generator on a Killing spinor $(\psi^A, \psi_A)$ is given by
\begin{equation}\label{Holonomy}
 \lef \psi^A\,,\,\psi_A \rig \;\mapsto\; -\, 2\, \alpha^2\, \lef B^{A}_{\ph{D}B}\,\,\psi^B \,,\,
    -\,\psi_B\,B^{B}_{\ph{B}A} \rig \,.
\end{equation}
Thus, it can be promptly recognized that if the eigenvalue $\alpha$ vanishes then the Killing spinor is left invariant by the action of the holonomy group. Since the non-trivial representations of the group $SO(p,6-p)$ have no fixed points, it follows that the Riemannian holonomy group must be special in such a case, namely the holonomy is a proper subgroup of $SO(p,6-p)$. Nevertheless, if $\alpha$ is non-vanishing then the holonomy group is not necessarily special. These features are not exclusive of six dimensions. Indeed, in any dimension the existence of a covariantly constant spinor reduces de holonomy while the existence of Killing spinors with non-vanishing eigenvalue generally is not associated with the reduction of the Riemannian holonomy. For instance, the $n$-dimensional sphere $S^n$ has several Killing spinors with non-vanishing eigenvalue but its holonomy group is the full $SO(n)$. Although no exceptional result about the holonomy has been obtained in this paragraph, it is worth pointing out that Eq. (\ref{Holonomy}) can be of great value on the investigation of the holonomy of six-dimensional manifolds admitting Killing spinors. This, in turn, is of relevance to determine the possible number unbroken supersymmetries on spontaneous compactifications \cite{Freund}.


\section{Integrating the Killing Spinor Equation}\label{Sec.Integration}

The aim of the present section is to integrate the Killing spinor equation in a six-dimensional manifold of Lorentzian signature. We shall not consider the quite special case of a covariantly constant spinor, namely we will assume that the eigenvalue $\alpha$ is non-vanishing. In particular, this assumption implies that the spinor is non-chiral. One of the reasons for omitting the case $\alpha=0$ is that such simpler case have already been quite scrutinized before \cite{ParallelSpinors-Wang,ParallelSpinors-Baum,KillingSpinorsHolonomy}. However, it is worth pointing out the interesting fact that, in manifolds with Euclidean metrics, it is possible to rescale the metric in such a way that a Killing spinor with imaginary eigenvalue\footnote{The sign conventions adopted here for the Clifford algebra, see Eq. (\ref{CliffordAlgebra}), are different from the ones adopted in \cite{Bar,O'Farrill}. Therefore, what here is called an imaginary eigenvalue, actually means a real eigenvalue in those references.} corresponds to a parallel spinor on the cone over the original manifold \cite{Bar,O'Farrill}. Therefore, in such cases, the study of a Killing spinor with non-zero eigenvalue reduces to the analysis of a covariantly constant spinor in a manifold with one more dimension. But, here, we will work in the Lorentzian signature and the eigenvalue $\alpha$ can be either real or imaginary, so that this approach cannot be used.

In what follows, we shall look for solutions for the equations
\begin{equation}\label{KillingSpinor6D3}
 \nabla_{AB} \, \psi^C \eq 2\,\alpha\,\delta^C_{\;[A}\,\psi_{B]} \quad\textrm{ and }\quad \nabla_{AB} \, \psi_C \eq -\,\alpha\, \varepsilon_{ABCD}\,\psi^D\,,
\end{equation}
namely the Killing spinor equation, in the Lorentzian signature when $\alpha\neq 0$. Our choice of signature is implemented by the use of the reality conditions (\ref{Reality_Vectors}) and (\ref{Conjugation}). The first advantage of using the index formalism adopted here is that one can, straight from the beginning, simplify the algebraic form of the spinor $(\psi^A, \psi_A)$ by means of using Lorentz transformations to properly align the spinorial  frame. Indeed, in Eqs. (\ref{AlgebraicSimp1}) and (\ref{AlgebraicSimp2}) of appendix \ref{AppendixLorentz} it has been shown that if $(\psi^A, \psi_A)$ is non-chiral spinor then one can always manage to choose the spinorial frame in such a way that this spinor field assumes one of the following algebraic forms:
\begin{equation}\label{StandardForms}
   \left\{
     \begin{array}{cl}
       \bl{\psi} \eq \lef \bl{\chi}_1 \,,\, f \,\bl{\zeta}^4 \rig  &\rightarrow\quad  \textrm{If $\psi^A\psi_A$ and $\psi^A\overline{\psi}_A$ both vanish} \\
                \\
 \bl{\psi} \eq \lef \bl{\chi}_1 \,,\, f_1 \,\bl{\zeta}^1 +  f_2 \,\bl{\zeta}^2  \rig    &\rightarrow\quad  \textrm{If at least one of the scalars $\psi^A\psi_A$ and $\psi^A\overline{\psi}_A$ is non-vanishing}\,,\\
     \end{array}
   \right.
\end{equation}
where $f$ is a real function while $f_1$ and $f_2$ are complex functions. Furthermore, we can check that we must assume $f_2 \neq 0$. Indeed, contracting Eq. (\ref{KillingSpinor6D3}) with $\psi^A \,\overline{\psi}^B$ we find that
$$ \psi^A \,\overline{\psi}^B\,\nabla_{AB} \, \psi^C \eq \alpha\, \lef \psi^C\, \overline{\psi}^A\,\psi_A \me  \overline{\psi}^C \,\psi^A\, \psi_A\rig \quad\textrm{ and }
\quad \psi^A \,\overline{\psi}^B\,\nabla_{AB} \, \psi_C \eq 0  \,.$$
Then, by means of contracting the first of these relations with $\overline{\psi}_C$ and using the charge conjugated version of the second identity above, we conclude that
$$  \psi^A \,\overline{\psi}^B\,\nabla_{AB} \lef \psi^C \,\overline{\psi}_C \rig  \eq
 \alpha\, \lef \overline{\psi}_C \,\psi^C\, \overline{\psi}^A\,\psi_A \me  \overline{\psi}_C \,\overline{\psi}^C \,\psi^A \, \psi_A\rig  \,.$$
Therefore,  if $ \psi^C \,\overline{\psi}_C$ is identically zero then $\psi^A  \psi_A$ must vanish, since we are assuming $\alpha \neq 0$. In terms of the second standard form in Eq. (\ref{StandardForms}), this means that if $f_2= 0$ then $f_1 =0$, which yields a chiral spinor and, therefore, requires
$\alpha = 0$. Thus, the condition $f_2 =0$ is inconsistent with the hypothesis $\alpha\neq 0$. Additionally, in accordance with Table \ref{TableSymmetryTensors}, it follows that the scalar $\psi^A\psi_A$ is constant, so that we can set $f_1$ equal to some constant. Moreover, from the same Table, it follows that $\psi^A\overline{\psi}_{A}+ \frac{\,\alpha^*}{\alpha} \overline{\psi}^{A}\psi_A$ is also constant, which means that the imaginary part of $\alpha f_2$ is constant. Therefore, we can always suppose that the Killing spinor has one of the following simple forms:
\begin{equation}\label{StandardForms2}
  \bl{\psi} \eq \lef \bl{\chi}_1 \,,\, f \,\bl{\zeta}^4   \rig  \quad  \textrm{ or } \quad
\bl{\psi} \eq \lef \bl{\chi}_1 \,,\, c_1 \,\bl{\zeta}^1 +  \frac{1}{\alpha}(h + i\,c_2) \,\bl{\zeta}^2  \rig \,.
\end{equation}
Where $f$  and $h$ are real functions,  $c_2$ is a real constant and $c_1$ is a complex constant. In addition, $h$ and $c_2$ cannot be simultaneously zero.  In the sequel, we shall try to integrate the Killing spinor equation for these two algebraic types of spinors. At this point, it is worth highlighting that,  for the Lorentzian signature, the study of these two possibilities is exhaustive.

\subsection{Killing spinors of the type $(\bl{\chi}_1 , f \bl{\zeta}^4)$}\label{Subsection-KStype1}

Let us start trying to integrate the Killing spinor equation for the first algebraic type in Eq. (\ref{StandardForms2}), namely let us impose that $\bl{\psi} \eq (\bl{\chi}_1 , f \bl{\zeta}^4)$ is a Killing spinor with eigenvalue $\alpha$. Then, its charge conjugate $\overline{\bl{\psi}} \eq (\bl{\chi}_2 , f \bl{\zeta}^3)$ is a Killing spinor with eigenvalue $\alpha^*$, where it has been used the fact that $f$ is a real function. Therefore, the components of the Killing spinor and its conjugate are given by
\begin{equation}\label{Components_Type1}
\psi^A = \delta^A_{\;1} \;\;\;, \quad  \psi_A = f\,\delta^{\,4}_{\;A}  \;\;\;  \textrm{and} \;\;\;  \overline{\psi}^A = \delta^A_{\;2} \;\;\;, \quad
 \overline{\psi}_A = f\,\delta^{\,3}_{\;A}  \,.
\end{equation}
Since, for this algebraic type of Killing spinor, all possible scalars that could be built using $\bl{\psi}$ and $\overline{\bl{\psi}}$ vanish, it follows that the geometric structures descendant from this Killing spinor have a null character. For instance, the Killing vector field constructed from these spinors have vanishing norm.  Inserting the components (\ref{Components_Type1}) in the integrability conditions (\ref{IntegrabCondSpin1}) and (\ref{IntegrabCondSpin2}), one verifies that the only components of the Weyl tensor that can be different from zero are the 9 components $\Psi^{ij}_{\ph{ij}pq}$, where $i,j\in \{1,2\}$ and $p,q\in \{3,4\}$. According to the CMPP (Coley, Milson, Pravda and Pravdov\'{a}) classification \cite{CMPP1,CMPP2}, which is a generalization of the Petrov classification for higher dimensions, this implies that the algebraic type of the Weyl tensor is either $N$ or $O$ \cite{Spin6D}, with $\bl{e}_1$ being the repeated principal null direction. This could be anticipated from the fact that Weyl tensors of type $N$ are generally associated with null structures, as exemplified by the $pp$-wave spacetime. Analogously, the integrability conditions imply that the unique component of the object $\Phi^{AB}_{\ph{AB}CD}$ that can be non-vanishing is $\Phi^{12}_{\ph{12}34}$, which means that the only possible obstruction for the spacetime to be Einstein comes from the component
$$  S_{44} \eq  \lef R_{\mu\nu}  \me  \frac{1}{6}\,R\,g_{\mu\nu} \rig e_4^{\;\mu}\,e_4^{\;\nu}  $$
of the traceless part of the Ricci tensor. Thus, a spacetime admitting a Killing spinor of the first type displayed in Eq. (\ref{StandardForms2}) is an Einstein manifold if, and only if, $S_{44}$ is zero.

Inserting these components in the Killing spinor equation (\ref{KillingSpinor6D3}) and its conjugated version, and then using Eq. (\ref{Cov Deriv Spinor6D}), we are led to the following restrictions over the spinorial connection:
\begin{equation}\label{Connection1}
  \left.
     \begin{array}{ll}
      (\Omega_{AB})^C_{\ph{C}1} &\eq 2\,\alpha\,f\,\delta^C_{\;[A}\,\delta^{\,4}_{\;B]}\;\;, \\
\\
       (\Omega_{AB})^C_{\ph{C}2} &\eq 2\,\alpha^*\,f\,\delta^C_{\;[A}\,\delta^{\,3}_{\;B]}\;\;, \\
     \end{array}
   \right.  \quad
\left.
     \begin{array}{ll}
            (\Omega_{AB})^4_{\ph{4}C} &\eq  \delta^4_{\;C}\,\frac{1}{f}\,\partial_{AB} f \ma  \frac{\alpha}{f} \, \varepsilon_{ABC\,1} \;\;, \\
\\
       (\Omega_{AB})^3_{\ph{4}C} &\eq  \delta^3_{\;C}\,\frac{1}{f}\,\partial_{AB} f \ma  \frac{\alpha^*}{f} \, \varepsilon_{ABC\,2} \;.
     \end{array}
   \right.
\end{equation}
In particular, the first and the last of the above relations imply:
\begin{equation}\label{Constraintf}
(\Omega_{34})^3_{\ph{3}1}  \eq \alpha\,f   \quad \textrm{and}  \quad   (\Omega_{34})^3_{\ph{3}1}  \eq \frac{\alpha^*}{f}  \quad  \Rightarrow \quad
f^2 \eq \frac{\alpha^*}{\alpha} \,.
\end{equation}
Since $f$ is a real function, it follows from the latter equation that $\alpha$ must be real. Which, due to (\ref{AlphaR}), means that the Ricci scalar is negative. Therefore, we have proved the following interesting result: \emph{In a six-dimensional Lorentzian manifold, the Killing spinor equation can admit a solution $(\psi^A, \psi_A)$ such that both $\psi^A \psi_A$ and $\overline{\psi}^A \psi_A$ vanish only if the curvature scalar is negative.} For instance, asymptotically de Sitter spacetimes cannot admit this type Killing spinor.  It is worth stressing the central role of the index notation on the proof the latter statement, as this result would be much harder to obtain otherwise. Thus, we can assume that $\alpha$ is real in Eq. (\ref{Connection1}). In this case, Eq. (\ref{Constraintf}) implies that $f = \pm 1$. By means of a Lorentz transformation, we can always make the sign of $f$ positive. Therefore, our Killing spinor might have the form $\bl{\psi} \eq (\bl{\chi}_1 ,  \bl{\zeta}^4)$ and its eigenvalue is real, so that Eq. (\ref{Connection1}) provides the following restrictions on the spinorial connection:
\begin{equation}\label{Connection2}
  \left.
     \begin{array}{ll}
      (\Omega_{AB})^C_{\ph{C}1} \eq 2\,\alpha\,\delta^C_{\;[A}\,\delta^{\,4}_{\;B]}\;\;,\\
\\
       (\Omega_{AB})^C_{\ph{C}2} \eq 2\,\alpha\,\delta^C_{\;[A}\,\delta^{\,3}_{\;B]}\;\;,\\
     \end{array}
   \right.  \quad
\left.
     \begin{array}{ll}
        (\Omega_{AB})^4_{\ph{4}C} \eq   \alpha \, \varepsilon_{ABC\,1} \;\;,  \\
\\
         (\Omega_{AB})^3_{\ph{4}C} \eq  \alpha \, \varepsilon_{ABC\,2} \;.
     \end{array}
   \right.
\end{equation}
Apart from the obvious skew-symmetry  $(\Omega_{AB})^C_{\ph{C}D}= - (\Omega_{BA})^C_{\ph{C}D}$, these equations impose that the only components of the spinorial connection that can be non-vanishing  are:
\begin{equation}\label{Connection3}
  \left\{
    \begin{array}{ll}
     (\Omega_{14})^1_{\ph{c}1} \eq  (\Omega_{41})^3_{\ph{c}3} \eq (\Omega_{24})^2_{\ph{c}1}
\eq  (\Omega_{34})^3_{\ph{c}1} \eq (\Omega_{13})^1_{\ph{c}2} \eq \alpha  \,,\\
\\
       (\Omega_{43})^4_{\ph{c}2} \eq  (\Omega_{23})^2_{\ph{c}2} \eq (\Omega_{32})^4_{\ph{c}4}
\eq  (\Omega_{24})^4_{\ph{c}3} \eq (\Omega_{13})^3_{\ph{c}4} \eq \alpha \,, \\
\\
     (\Omega_{AB})^1_{\ph{c}3} \;,\; \;  (\Omega_{AB})^1_{\ph{c}4} \;,\; \; (\Omega_{AB})^2_{\ph{c}3} \;,\;\;  (\Omega_{AB})^2_{\ph{c}4} \,.
    \end{array}
  \right.
\end{equation}
Where the components in the last line of (\ref{Connection3}) are not constrained by Eq. (\ref{Connection2}), so that each of these components are, at this stage, arbitrary. Thus, integrating the Killing spinor equation for a spinor of the form $\bl{\psi} \eq (\bl{\chi}_1 ,  \bl{\zeta}^4)$ amounts to finding the most general metric such that the non-vanishing components of the spinorial connection are the ones displayed in Eq. (\ref{Connection3}).  By means of the results presented in appendix \ref{AppendixConnection}, it follows that Eq. (\ref{Connection3}) is tantamount to imposing that the only components of the tangent bundle connection $\omega_{ab}^{\ph{ab}c}$ that can be different from zero are the following:
\begin{equation}\label{Connection4}
  \left\{
    \begin{array}{ll}
 \omega_{31}^{\ph{31}1} \eq  - \omega_{34}^{\ph{34}4} \eq \omega_{33}^{\ph{33}3} \eq - \omega_{36}^{\ph{36}6} \eq
 \omega_{61}^{\ph{61}1}  \eq  - \omega_{64}^{\ph{64}4} \eq \omega_{66}^{\ph{66}6} \eq - \omega_{63}^{\ph{63}3}  \eq
  \omega_{26}^{\ph{26}2} \eq - \omega_{25}^{\ph{25}3}  \eq  \alpha  \,,  \\
\\
       \omega_{23}^{\ph{23}2} \eq  - \omega_{25}^{\ph{25}6} \eq
\omega_{43}^{\ph{43}4} \eq - \omega_{41}^{\ph{41}6} \eq  \omega_{46}^{\ph{46}4}  \eq  - \omega_{41}^{\ph{41}3} \eq
\omega_{53}^{\ph{53}5} \eq - \omega_{52}^{\ph{52}6}  \eq   \omega_{56}^{\ph{56}5} \eq - \omega_{52}^{\ph{52}3}  \eq  \alpha    \,, \\
\\
    \omega_{a4}^{\ph{a4}3} \eq -   \omega_{a6}^{\ph{a6}1} \;,\; \; \omega_{a4}^{\ph{a4}2}  \eq -  \omega_{a5}^{\ph{a5}1} \;,\; \;
 \omega_{a4}^{\ph{a4}5}  \eq -   \omega_{a2}^{\ph{a2}1} \;,\; \; \omega_{a4}^{\ph{a4}6}  \eq -   \omega_{a3}^{\ph{a3}1} \,,
    \end{array}
  \right.
\end{equation}
where the skew-symmetry $\omega_{abc} = -\omega_{acb}$ have been used. For instance, the latter property implies  that $\omega_{a3}^{\ph{a3}2}= -\omega_{a5}^{\ph{a5}6}$. Recall that,  $\omega_{abc}$ is defined by $\omega_{abc} \eq \omega_{ab}^{\ph{ab}d}g_{dc} \eq\bl{g}(\nabla_a e_b, e_c)$, which comes from Eq. (\ref{TangentConnection}).

Now let us try to find a metric whose Levi-Civita connection obey the constraints shown in Eq. (\ref{Connection4}). As a first step toward  this goal, let us note that, since $\psi^A\psi_A$ vanishes, it follows that the chiral spinors $\psi^A$ and $\psi_A$ are pure spinors whose associated maximally isotropic distributions are integrable. Indeed, the vector fields that annihilate the chiral spinor $\psi^A$ under the  Clifford action are the ones that have the form
$V^{AB} \eq \psi^{[A} \phi^{B]}$ for some spinor $\phi^B$, as can be easily checked using (\ref{CliffordAction6D}). Analogously, the vector fields that annihilate the negative chirality spinor $\psi_A$ are the ones that have the form $V_{AB} \eq \psi_{[A} \xi_{B]}$ for some spinor $\xi_B$. Now, using the Killing spinor equation (\ref{KilliingSpinor6D1}) satisfied by $(\psi^A, \psi_A)$, it follows that
$$ \psi^{A} \phi^{B}\,\nabla_{AB}\,\psi^C \,\propto\,\psi^C \quad \textrm{and} \quad  \psi_{A} \xi_{B}\,\nabla^{AB}\,\psi_C \,\propto\,\psi_C \,,  $$
which means that the maximally isotropic distributions generated by $\psi^A$ and $\psi_A$ are both integrable \cite{Batista-PureSubspace}. These distributions are the ones generated by $\{\bl{e}_1,\bl{e}_2, \bl{e}_3 \}$ and  $\{\bl{e}_1,\bl{e}_2, \bl{e}_6 \}$ respectively. Analogously, since $(\overline{\psi}^A, \overline{\psi}_A)$ is also a Killing spinor and $\overline{\psi}^A \overline{\psi}_A=0$, it follows that the spinors $\bl{\chi}_2$ and $\bl{\zeta}^3$ generate integrable distributions, which are spanned by $\{\bl{e}_1,\bl{e}_5, \bl{e}_6 \}$ and  $\{\bl{e}_1,\bl{e}_3, \bl{e}_5 \}$ respectively. The fact that these four totally null distributions are integrable can also be verified directly from Eq. (\ref{Connection4}). Now, since the intersection of two integrable distributions is also an integrable distribution, we conclude that the distributions spanned by $\{\bl{e}_1,\bl{e}_2\}$, $\{\bl{e}_1,\bl{e}_3\}$, $\{\bl{e}_1,\bl{e}_5\}$ and $\{\bl{e}_1,\bl{e}_6\}$ are all integrable.

Inasmuch as $\bl{e}_1$ is a real vector field in the Lorentzian signature, we can always introduce a real coordinate $u$ to be the parameter along the orbits of $\bl{e}_1$, so that we can set
$$ \bl{e}_1 \eq \partial_u \,.$$
Then, due to the fact that the distribution generated by $\{\bl{e}_1,\bl{e}_2\}$ is integrable, it follows from the Frobenius theorem that there exists a coordinate $z$ such that
$$  \bl{e}_2 \eq U_2  \, \partial_u \ma Z_2\, \partial_z\,, $$
where $U_2$ and $Z_2$ are functions. Note that $U_2$, $Z$ and $z$ are generally complex, since, in the Lorentzian signature, $\bl{e}_2$ is a complex vector field. Likewise, the fact that  the distribution generated by $\{\bl{e}_1,\bl{e}_3\}$ is integrable implies that there exists a complex coordinate $w$ such that
$$  \bl{e}_3 \eq U_3  \, \partial_u \ma W_3\, \partial_w\,, $$
with $U_3$ and $W_3$ being complex functions. Moreover, due to the reality conditions of Eq. (\ref{Reality_Vectors}), we conclude that
$$ \bl{e}_5 \eq U_2^* \, \partial_u \ma Z_2^*\, \partial_{z^*}  \quad \textrm{ and} \quad
\bl{e}_6 \eq U_3^*  \, \partial_u \ma W_3^*\, \partial_{w^*} \,. $$
Finally, $\bl{e}_4$ can have the following general form:
$$  \bl{e}_4 \eq   V_4  \, \partial_v \ma U_4  \, \partial_u \ma   Z_4\, \partial_z  \ma W_4\, \partial_w \ma Z_4^*\, \partial_{z^*}  \ma W_4^*\, \partial_{w^*}\,, $$
where $v$ is a real coordinate, $V_4$ and $U_4$ are real functions whereas $Z_4$ and $W_4$ are complex functions. In principle, the functions that appear as coefficients of the above vector fields can depend on all six coordinates $\{u,v,z,z^*,w,w^*\}$. However, according to Table \ref{TableSymmetryTensors}, the vector field $\psi^{[A}\overline{\psi}^{B]} \me \frac{1}{2} \varepsilon^{ABCD} \psi_C\overline{\psi}_D$ is a Killing vector field. Since the latter vector field is $2\, \chi_1^{\;[A}\chi_2^{\;B]} \,\sim\, 2\bl{e}_1 \eq 2\,\partial_u$, it follows that the above functions cannot depend on the coordinate $u$. Furthermore, computing the Lie brackets $[\bl{e}_2,\bl{e}_3]$, $[\bl{e}_2,\bl{e}_4]$,  $[\bl{e}_2,\bl{e}_5]$ and using that
$$ [\bl{e}_a,\bl{e}_b] \eq (\omega_{ab}^{\ph{ab}c}\me  \omega_{ba}^{\ph{ab}c})\,\bl{e}_c $$
along with Eq. (\ref{Connection4}), we obtain that the functions $W_3$ and $V_4$ do not depend on the coordinates $z$ and $z^*$, while $Z_2$ cannot depend on $z^*$. So, it is tempting to look for solutions that do not depend on $z$ and $z^*$. After imposing (\ref{Connection4}) and making some minor simplifying assumptions along the integration process, we have found that a particular solution of this type is given by the following functions:
\begin{equation*}
  \left\{
    \begin{array}{ll}
       U_2 \eq 0 \;,\quad Z_2 \eq i\,\frac{(w^2 \ma w^{*\, 2})}{h(v)} \;, \\
\\
 U_3 \eq \alpha\, \frac{(w^2 \ma w^{*\, 2})}{w}\,\left[ \,F(v,y_1 )\ma  i\, G(v, y_2) \,\right] \;,\quad   W_3  \eq -\alpha\,\frac{(w^2 \ma w^{*\, 2})}{2\,w} \;, \\
\\
 U_4 \eq  (w^2 \ma w^{*\, 2})^4\,g(v) \;,\quad   V_4 \eq  2\,(w^2 \ma w^{*\, 2})^2   \;,\quad Z_4 \eq 0 \;,\quad W_4 \eq 0 \,.
    \end{array}
  \right.
\end{equation*}
Where $h(v)$, $g(v)$, $F(v,y_1)$, $G(v,y_2)$ are arbitrary real functions of their arguments and $y_1$ is the real part of the complex coordinate $w$ whereas $y_2$ is the imaginary part of $w$,
$$ y_1 \eq \frac{1}{2}\lef w \ma w^* \rig \quad, \quad  y_2 \eq \frac{1}{2\,i}\lef w \me w^* \rig \,. $$
Once we have these functions at hand, one can find the coordinate form of the metric. Indeed, due to the inner products (\ref{NullFrameInnerP}) it follows that the inverse of the metric is given by
$$ \bl{g}^{-1} \eq   2\lef \bl{e}_1\otimes \bl{e}_4 \ma \bl{e}_4\otimes \bl{e}_1 \ma     \bl{e}_2\otimes \bl{e}_5 \ma \bl{e}_5\otimes \bl{e}_2 \ma     \bl{e}_3\otimes \bl{e}_6 \ma \bl{e}_6\otimes \bl{e}_3  \rig \,.  $$
Then, inserting the coordinate form of these vector fields and computing the inverse, we end up with the following line element:
\begin{align}
  ds^2 \eq \frac{-1}{4}\,g\,dv^2 \ma  \frac{1}{ (w^2 \ma w^{*\, 2})^2 }  &\,\bigg[ \,\frac{1}{2}\,du dv \ma h^2\,dz\,dz^* \ma \frac{4\,w\,w^*}{\alpha^2} \,dwdw^*    \nonumber\\
  & \quad      \ma  (F + i G)\,dv dw \ma  (F - i G)\,dv dw^*  \,\bigg]\,. \label{Metric1}
\end{align}
Now, let us discuss about the symmetry tensors of this spacetime. As anticipated, the Killing vector field generated by the Killing  spinor, using the results of Table \ref{TableSymmetryTensors}, is given by $\bl{e}_1$. Besides, it is obvious that $\partial_z$ and  $\partial_{z^*}$ are also Killing vectors.  The closed conformal Killing vector presented in Table  \ref{TableSymmetryTensors} is identically zero for the Killing spinor $(\bl{\chi}_1 , \bl{\zeta}^4)$. The KY tensor of rank 2, given by $ \psi^A\psi_B - \frac{1}{4}(\psi^C  \psi_C)\delta^A_B$, and its complex conjugate are given respectively by
$$ \bl{Y}_1 \eq \bl{e}_1 \wedge \bl{e}_2  \quad \textrm{ and }\quad \overline{\bl{Y}_1} \eq \bl{e}_1 \wedge \bl{e}_5 \,,$$
where $\wedge$ denotes the skew-symmetric tensor product of the vector fields and multiplicative constants have been omitted. The remaining KY bivector and the closed conformal Killing-Yano bivector, presented in the bottom half of Table \ref{TableSymmetryTensors}, are respectively given by
$$ \bl{Y}_2 \eq \bl{e}_1 \wedge (\bl{e}_3 -  \bl{e}_6)  \quad \textrm{ and }\quad
\bl{H} \eq \bl{e}_1 \wedge (\bl{e}_3 +  \bl{e}_6)   \,.$$
Further,  the Killing-Yano 3-vectors extracted from the same table are given by
$$ \bl{T}_1 \eq \bl{e}_1 \wedge \bl{e}_2 \wedge (\bl{e}_3 - \bl{e}_6)  \quad , \quad
\overline{\bl{T}_1} \eq  \bl{e}_1 \wedge \bl{e}_5 \wedge (\bl{e}_6 - \bl{e}_3) \quad \textrm{ and }\quad
 \bl{T}_2  \eq  \bl{e}_1 \wedge \bl{e}_2 \wedge \bl{e}_5 \,.  $$
The Hodge dual of these three Killing-Yano 3-vectors are CCKY 3-vectors.

It is worth stressing that the metric (\ref{Metric1}) is not an Einstein metric. This is not in contradiction with the existence of a Killing spinor, since only in Euclidean signature does the existence of a solution for the Killing spinor equation requires the manifold to be Einstein, as emphasized before.  In the particular case in which the functions $F$ and $G$ do not depend on $v$, the Einstein condition can be easily solved and is given by choosing
\begin{equation}\label{SolutionEinstein1}
  g(v) \eq -\,\frac{2}{3\,\alpha^2\,h(v)}\, \frac{d^2}{dv^2}h(v) \;\;,\quad F \eq F(y_1) \;\;\textrm{ and }\quad G \eq G(y_2) \,,
\end{equation}
where $h(v)$, $F(y_1)$ and $G(y_2)$ remain being arbitrary real functions of their arguments.  This solution for Einstein's equation in the presence of a negative cosmological constant is not conformally flat and, therefore, comprises a non-trivial solution. By means of computing the Riemann tensor of this metric, it has been checked that the Riemannian holonomy group of this spacetime is $SO(1,5)$. Concerning the isometry group, it follows that in general this solution admits only the three obvious Killing vector fields $\partial_u$,  $\partial_z$ and $\partial_{z^*}$. However, in the particular case in which $h(v)^2$ is a quadratic function of $v$ it follows that one further Killing vector field shows up, whose expression is given by:
$$  h^2\,\,\partial_v + \frac{1}{4}\, w \,\frac{d\,h^2}{dv} \, \partial_w \ma  \frac{1}{4}\, w^* \,\frac{d\,h^2}{dv} \, \partial_{w^*}
    - \,\left[\,  \frac{|w|^4}{\alpha^2}\,\frac{d^2\,h^2}{dv^2} \ma \frac{d\,h^2}{dv} \lef  y_1\,F(y_1) \me y_2\,G(y_2)   \rig \,\right]\,   \partial_u  \,.$$
Where, again,  $y_1$ and $y_2$  are  the real and imaginary parts of the coordinate $w$, respectively. Therefore, in general, the spacetime defined by Eqs. (\ref{Metric1}) and (\ref{SolutionEinstein1}) has a 3-dimensional abelian isometry group. However, in the special case in which $h^2$ is a quadratic function of $v$, it turns out that the isometry group is enhanced and becomes a 4-dimensional abelian group.


\subsection{Killing spinors of the type $( \bl{\chi}_1 \, , \, c \,\bl{\zeta}^1 +   f\, \bl{\zeta}^2)$}


Now it is time to analyse the Killing spinor equation for spinors whose algebraic form can be chosen to be $\bl{\psi}= (\bl{\chi}_1 \,,\, c \,\bl{\zeta}^1 +   f\, \bl{\zeta}^2)$. At the beginning of the present section, we have already argued that in order for $\bl{\psi}$ to be a Killing spinor with non-vanishing eigenvalue, $c$ must be a constant and $f$ must take the form $f=\frac{1}{\alpha}(h + i\,c_2)$, where $h$ is a real function, $c_2$ is a real constant, and  $|f|^2 \neq 0$. As a consequence of the latter constraint, one can check that the integrability conditions (\ref{IntegrabCondSpin1}) and (\ref{IntegrabCondSpin2}) for $\bl{\psi}$ and its charge conjugate,  $\overline{\bl{\psi}}= (\bl{\chi}_2 \,,\, c^* \,\bl{\zeta}^2 -   f^*\, \bl{\zeta}^1)$, yield the following constraints on the curvature
$$  \Psi^{AB}_{\ph{AB}Cj} \eq     \Phi^{AB}_{\ph{AB}Cj} \eq 0 \eq \Psi^{Aj}_{\ph{A1}CD}   \eq  \Phi^{Aj}_{\ph{A2}CD} \,,$$
where $j\in\{1,2\}$ and $A,\,B,\,C,\,D\,\in\{1,2,3,4\}$. In particular, this means that the only component of the traceless part of the Ricci tensor that could be different from zero is $\Phi^{34}_{\ph{34}34}$. However, due to the traceless condition (\ref{Algeb.Curvature}), it follows that the latter component must also vanish. Therefore, we arrive at the interesting conclusion that \emph{a six-dimensional Lorentzian manifold possessing a Killing spinor $\psi=(\psi^A,\psi_A)$ such that either $\psi^A\psi_A$ or $\psi^A\overline{\psi}_A$ is non-vanishing, is, necessarily, an Einstein spacetime}. Regarding the Weyl tensor, the only components that can be non-vanishing are $\Psi^{pq}_{\ph{pq}p'q'}$, where $p,q,p',q' \in\{3,4\}$. Thus, according to the CMPP classification \cite{CMPP1,CMPP2}, the algebraic type of the Weyl tensor is either $D$ or $O$, with $\bl{e}_1$ and $\bl{e}_4$ being repeated principal null directions \cite{Spin6D}. Type $D$ spacetimes are quite special from the physical point of view, since all spacetimes in the Kerr-NUT-($A$)$dS$ family of black holes, in arbitrary dimensions, have this algebraic type in the CMPP classification \cite{CMPP2}

Particularly, combining these results with the ones obtained in the previous subsection, it follows that in order for a spacetime to admit both algebraic types of Killing spinors its CMPP classification must be $O$, namely the Weyl tensor vanishes, it must be an Einstein manifold, and its Ricci scalar must be negative. Therefore, the only spacetime that could possibly admit both algebraic types of Killing spinors is Anti-de Sitter ($AdS$). Indeed, at the end of this subsection it is shown that this possibility is realized, namely $AdS_6$ spacetime admits both types of Killing spinors.

Inserting the spinor $\bl{\psi}= (\bl{\chi}_1 \,,\, c \,\bl{\zeta}^1 +   f\, \bl{\zeta}^2)$ into the Killing spinor equation (\ref{KillingSpinor6D3}) as well as its charge conjugate, which is a Killing spinor with eigenvalue $\alpha^*$, lead to the following constraints over the spinorial connection:
\begin{equation*}\label{Connection1-2}
  \left.
     \begin{array}{ll}
      (\Omega_{AB})^C_{\ph{C}1} &\eq 2\,\alpha\,c\,\delta^C_{\;[A}\,\delta^{\,1}_{\;B]} \ma 2\,\alpha\,f\,\delta^C_{\;[A}\,\delta^{\,2}_{\;B]}\;\;, \\
\\
    (\Omega_{AB})^C_{\ph{C}2} &\eq 2\,\alpha^*\,c^*\,\delta^C_{\;[A}\,\delta^{\,2}_{\;B]} \me 2\,\alpha^*\,f^* \,\delta^C_{\;[A}\,\delta^{\,1}_{\;B]}\;\;, \\
     \end{array}
   \right.  \quad
\left.
     \begin{array}{ll}
 c\,(\Omega_{AB})^1_{\ph{4}C} \ma  f\,(\Omega_{AB})^2_{\ph{4}C} & \eq  \alpha\, \varepsilon_{ABC\,1}\ma \delta^2_{\;C}\,\partial_{AB} f  \;\;, \\
\\
 c^*\,(\Omega_{AB})^2_{\ph{4}C} \me  f^*\,(\Omega_{AB})^1_{\ph{1}C} & \eq
\alpha^*\, \varepsilon_{ABC\,2} \me \delta^1_{\;C}\,\partial_{AB} f^* \;.
     \end{array}
   \right.
\end{equation*}
Manipulating these equations, we eventually conclude that the only components of the spinorial connection $(\Omega_{AB})^C_{\ph{C}D}$ that can be different from zero, apart from the skew-symmetry in the indices $AB$, are the following ones:
\begin{equation}\label{Connection5}
  \left\{
    \begin{array}{ll}
     (\Omega_{12})^1_{\ph{c}1} \eq  (\Omega_{32})^3_{\ph{c}1} \eq (\Omega_{42})^4_{\ph{c}1} \eq \alpha\, f \quad,\quad
     (\Omega_{12})^2_{\ph{c}2} \eq (\Omega_{13})^3_{\ph{c}2} \eq (\Omega_{14})^4_{\ph{c}2} \eq \alpha^*\, f^* \;, \\
\\
      (\Omega_{21})^2_{\ph{c}1} \eq  (\Omega_{31})^3_{\ph{c}1} \eq (\Omega_{41})^4_{\ph{c}1} \eq \alpha\, c \quad,\quad
     (\Omega_{12})^1_{\ph{c}2} \eq (\Omega_{32})^3_{\ph{c}2} \eq (\Omega_{42})^4_{\ph{c}2} \eq \alpha^*\, c^* \;,  \\
\\
     (\Omega_{24})^2_{\ph{c}3} \eq  (\Omega_{32})^2_{\ph{c}4} \eq \frac{\alpha \,f^* }{|c|^2 \ma |f|^2 } \quad,\quad
     (\Omega_{14})^1_{\ph{c}3} \eq  (\Omega_{31})^1_{\ph{c}4} \eq \frac{\alpha^* \,f }{|c|^2 \ma |f|^2 } \;,  \\
\\
(\Omega_{24})^1_{\ph{c}3} \eq  (\Omega_{32})^1_{\ph{c}4} \eq \frac{\alpha \,c^* }{|c|^2 \ma |f|^2 } \quad,\quad
     (\Omega_{13})^2_{\ph{c}4} \eq  (\Omega_{41})^2_{\ph{c}3} \eq \frac{\alpha^* \,c }{|c|^2 \ma |f|^2 } \;,  \\
\\
     (\Omega_{AB})^3_{\ph{c}3} \;,\; \;  (\Omega_{AB})^3_{\ph{c}4} \;,\; \; (\Omega_{AB})^4_{\ph{c}3} \;,\;\;  (\Omega_{AB})^4_{\ph{c}4}\;,
    \end{array}
  \right.
\end{equation}
where the components in the last line of the above list are arbitrary. In addition, we find that the function $f$ must satisfy the following differential constraints:
\begin{equation}\label{Dh}
  \partial_1 f \eq \alpha^* (|c|^2 \ma |f|^2 ) \quad, \quad  \partial_2 f \eq 0  \quad, \quad  \partial_3 f\eq 0 \quad, \quad  \partial_4  f \eq \alpha \quad, \quad  \partial_5 f \eq 0 \quad, \quad  \partial_6 f \eq 0 \,,
\end{equation}
where the indices subscribed on the partial derivatives are with respect to the null frame. For instance, $\partial_1 = e_1^{\;\mu}\partial_\mu $. Particularly, note that the function $f$ cannot be constant, otherwise we would have $\alpha = 0$ and the Killing spinor would be covariantly constant.

By means of Eq. (\ref{Connection5}) along with Eq. (\ref{Connection6D}), one can straightforwardly obtain the tangent bundle connection, $\omega_{ab}^{\ph{ab}c}$. For instance, in the particular case in which $c=0$, we have that the only components of the tangent bundle connection that can be different from zero are the following ones:
\begin{equation}\label{Connection6}
  \left\{
    \begin{array}{ll}
 \omega_{51}^{\ph{51}5}  \eq  - \omega_{52}^{\ph{52}4} \eq \omega_{61}^{\ph{61}6} \eq - \omega_{63}^{\ph{63}4}  \eq \alpha \, f
\quad , \quad
\omega_{21}^{\ph{21}2} \eq  - \omega_{25}^{\ph{25}4} \eq \omega_{31}^{\ph{31}3} \eq - \omega_{36}^{\ph{36}4} \eq \alpha^* \, f^* \quad,\\
 \\
 \omega_{54}^{\ph{54}5}  \eq  - \omega_{52}^{\ph{52}1} \eq \omega_{64}^{\ph{64}6} \eq - \omega_{63}^{\ph{63}1}  \eq \frac{\alpha}{f}
\quad , \quad
\omega_{24}^{\ph{24}2} \eq  - \omega_{25}^{\ph{52}1} \eq \omega_{34}^{\ph{34}3} \eq - \omega_{36}^{\ph{36}1} \eq \frac{\alpha^*}{f^*} \quad, \\
\\
    \omega_{12}^{\ph{12}2} \ma \omega_{13}^{\ph{13}3} \eq \alpha \, f \me \alpha^* \, f^* \quad , \quad
    \omega_{15}^{\ph{15}5} \ma \omega_{16}^{\ph{16}6} \eq  -\,(\alpha \,f \me \alpha^* \, f^*) \quad, \\
\\
    \omega_{11}^{\ph{11}1} \eq  - \omega_{14}^{\ph{14}4} \eq \alpha \, f \ma \alpha^* \, f^*  \quad , \quad
  \omega_{a1}^{\ph{a1}1} \eq -\, \omega_{a4}^{\ph{a4}4} \eq 0  \;\;  \textrm{if $a\neq 1$} \quad , \quad  \\
\\
\omega_{a2}^{\ph{a2}2} \ma \omega_{a3}^{\ph{a3}3} \eq 0\;\;  \textrm{if $a\neq 1$}\quad , \quad
    \omega_{a5}^{\ph{a5}5} \ma \omega_{a6}^{\ph{a6}6} \eq 0\;\; \textrm{if $a\neq 1$}\quad, \quad \\
\\    
    \omega_{a5}^{\ph{a5}6} \eq -   \omega_{a3}^{\ph{a3}2} \quad , \quad \omega_{a2}^{\ph{a2}3}  \eq -  \omega_{a6}^{\ph{a6}5} \quad .
     \end{array}
  \right.
\end{equation}
The other components of the tangent bundle connection not appearing in (\ref{Connection6}) must all be zero in order for the manifold to admit a Killing spinor of the form $\bl{\psi}= (\bl{\chi}_1 \,,   f\, \bl{\zeta}^2)$. Moreover, when $c= 0$, Eq. (\ref{Dh}) requires that the function $f$ satisfies the following differential conditions:
\begin{equation}\label{Dh2}
  \partial_1 f \eq \alpha^* \, |f|^2  \quad, \quad  \partial_2 f \eq 0  \quad, \quad  \partial_3 f \eq 0 \quad, \quad  \partial_4 f \eq \alpha \quad, \quad  \partial_5 f \eq 0 \quad, \quad  \partial_6 f \eq 0 \,.
\end{equation}
A particular solution of the constraints (\ref{Connection6}) and (\ref{Dh2}) is given by a spacetime whose line element is given by
\begin{equation}\label{AdS-metric}
  ds^2 \eq -\,f^2\,dt^2 \ma \frac{1}{f^2}\,dr^2 \ma r^2\,\lef dz \,dz^* \ma dw \,dw^* \rig \,,
\end{equation}
where $f=f(r) = 2\,\alpha \,r$, with the eigenvalue $\alpha$ assumed to be real. The coordinates $t$ and $r$ are real, while $z$ and $w$ are complex.  The latter metric is Einstein, with negative Ricci scalar, and conformally flat, i.e., it describes the Anti-de Sitter spacetime. In these coordinates, the null frame is given by
\begin{align}
 \bl{e}_1 &\eq \frac{1}{2}( f^2\,\partial_r \ma  \partial_t) \quad ,\quad  \bl{e}_2 \eq \frac{1}{r}\,\partial_z
\quad ,\quad  \bl{e}_3 \eq \frac{1}{r}\,\partial_w \;, \nonumber\\
 \bl{e}_4 &\eq \frac{1}{2}( \partial_r \me \frac{1}{f^2}\partial_t)
\quad ,\quad  \bl{e}_5 \eq \frac{1}{r}\,\partial_{z^*}  \quad ,\quad  \bl{e}_6 \eq \frac{1}{r}\,\partial_{w^*} \; \label{NullFrameAdS}.
\end{align}
By means of Table \ref{TableSymmetryTensors}, along with the fact that the Killing spinor has the form
$\bl{\psi}= (\bl{\chi}_1 \,,   f\, \bl{\zeta}^2)$, one can find the following Killing-Yano tensors of rank 1, 2 and 3:
$$  \bl{\mathcal{Y}}_1 \eq (\bl{e}_1 \me  f^2\,\bl{e}_4 ) \;\; , \quad \bl{\mathcal{Y}}_2 \eq f\,\bl{e}_2\wedge \bl{e}_3  \;\;  , \quad
\bl{\mathcal{Y}}_3 \eq f\,(\bl{e}_2\wedge \bl{e}_5 \ma  \bl{e}_3\wedge \bl{e}_6) \;\;
 , \quad \bl{\mathcal{Y}}_4 \eq  \frac{1}{f}\,\bl{\mathcal{Y}}_1\wedge\bl{\mathcal{Y}}_2 \;\; , \quad \bl{\mathcal{Y}}_5 \eq \frac{1}{f}\,\bl{\mathcal{Y}}_1\wedge\bl{\mathcal{Y}}_3\,, $$
where it has been used the fact that $\alpha$ was assumed to be real. Note that the Killing vector field $\bl{\mathcal{Y}}_1$ is just $\partial_t$. Using the same table, we also arrive at the closed conformal Killing-Yano tensors below:
$$  \bl{\mathcal{H}}_1 \eq (\bl{e}_1 \ma  f^2\,\bl{e}_4 ) \;\; , \quad \bl{\mathcal{H}}_2 \eq f\,\bl{e}_1\wedge \bl{e}_4  \;\;  , \quad
\bl{\mathcal{H}}_3 \eq \frac{1}{f}\,\bl{\mathcal{H}}_1\wedge\bl{\mathcal{Y}}_2  \,. $$

Although we have only worked out the symmetries associated with the Killing spinor $\bl{\psi}= (\bl{\chi}_1 \,,   f\, \bl{\zeta}^2)$, it is worth stressing that there are other Killing spinors in $AdS_6$ spacetime. Indeed, it is well-known that the maximally symmetric spaces also have the maximum number of Killing spinors, which is the number of components of a Dirac spinor \cite{AdS_spinors}. Thus, in the six-dimensional case, the $AdS$ spacetime with Ricci scalar given by $R= -120\alpha^2$ admits 8 independent Killing spinors with eigenvalue $\alpha$. Furthermore, there are 8 other spinors with eigenvalue $-\,\alpha$. Indeed, integrating the Killing equation  (\ref{KillingSpinor6D3}) in the $AdS_6$ background, using the coordinates (\ref{AdS-metric}) and the null frame (\ref{NullFrameAdS}), one can check that the most general Killing spinor with eigenvalue $\alpha$ is given by
\begin{equation}\label{KS-AdS}
  \bl{\psi}\eq \left(\, c_1\, \bl{\chi}_1  \ma c_2\, \bl{\chi}_2 \ma c_3\, f\,\bl{\chi}_3 \ma c_4\, f\,\bl{\chi}_4 \,\,,\,   -\,c_2\,  f\, \bl{\zeta}^1 \ma  c_1\,  f\, \bl{\zeta}^2  \me  c_4\, \bl{\zeta}^3  \ma  c_3\, \bl{\zeta}^4 \,\right)\,,
\end{equation}
where $c_1$, $c_2$, $c_3$ and  $c_4$ are arbitrary complex constants, so that the general solution is labeled by 8 real parameters. Regarding the most general solution with eigenvalue $-\,\alpha$, it can be obtained by acting with the chirality operator $\Upsilon$ on the previous spinor, which yields
\begin{equation}\label{KS-AdS2}
 \widetilde{\bl{\psi}} \eq \Upsilon\,\bl{\psi}\eq \left(\, c_1\, \bl{\chi}_1  \ma c_2\, \bl{\chi}_2 \ma c_3\, f\,\bl{\chi}_3 \ma c_4\, f\,\bl{\chi}_4 \,\,,\,   c_2\,  f\, \bl{\zeta}^1 \me  c_1\,  f\, \bl{\zeta}^2  \ma  c_4\, \bl{\zeta}^3  \me  c_3\, \bl{\zeta}^4 \,\right)\,.
\end{equation}
Computing the scalars that can be constructed from the general Killing spinor (\ref{KS-AdS}), one finds
$$ \psi^A\,\psi_A \eq 0 \quad, \quad
 \overline{\psi}^A\,\psi_A \eq -\, \psi^A\,\overline{\psi}_A \eq  f\lef |c_1|^2 \ma |c_2|^2 \me |c_3|^2 \me |c_4|^2 \rig \,. $$
Therefore, the Killing spinors of Eq. (\ref{KS-AdS}) such that $|c_1|^2 + |c_2|^2 = |c_3|^2 + |c_4|^2$ are of the algebraic type treated in Section \ref{Subsection-KStype1}, although in the basis adopted in (\ref{KS-AdS}) it is not possible to see that these Killing spinors have the form $(\bl{\chi}_1 , \bl{\zeta}^4)$. Thus, in the $AdS_6$ spacetime there are spinors of both algebraic types described in (\ref{StandardForms}).

Likewise, it is simple matter to integrate the Killing spinor equation in the de Sitter ($dS$) background. In the latter case the Ricci scalar is positive, so that the eigenvalue $\alpha$ must be imaginary. The metric of $dS_6$ spacetime with Ricci scalar $R=120|\alpha|^2 = - 120 \alpha^2$ can written  as follows:
\begin{equation*}\label{dS-metric}
  ds^2 \eq -\,|h|^2\,d\tau^2 \ma \frac{1}{|h|^2}\,dx^2 \ma \tau^2\,\lef dz \,dz^* \ma dw \,dw^* \rig \,,
\end{equation*}
where $h= h(\tau)= 2\,\alpha\,\tau$ with $\alpha$ being imaginary. Then, adopting the null frame
\begin{align*}
 \bl{e}_1 &\eq \frac{1}{2}( h^2\,\partial_\tau \me  \partial_x) \quad ,\quad  \bl{e}_2 \eq \frac{1}{\tau}\,\partial_z
\quad ,\quad  \bl{e}_3 \eq \frac{1}{\tau}\,\partial_w \;, \nonumber\\
 \bl{e}_4 &\eq \frac{1}{2}( \partial_\tau \ma \frac{1}{h^2}\partial_x)
\quad ,\quad  \bl{e}_5 \eq \frac{1}{\tau}\,\partial_{z^*}  \quad ,\quad  \bl{e}_6 \eq \frac{1}{\tau}\,\partial_{w^*} \; \label{NullFramedS},
\end{align*}
one can check that the most general solution for the Killing spinor equation with eigenvalue $\alpha$ is given by
\begin{equation}\label{KS-dS}
 \bl{\psi}\eq \left(\, b_1\, \bl{\chi}_1  \ma b_2\, \bl{\chi}_2 \ma b_3\, h\,\bl{\chi}_3 \ma b_4\, h\,\bl{\chi}_4 \,\,,\,   -\,b_2\,  h\, \bl{\zeta}^1 \ma  b_1\,  h\, \bl{\zeta}^2  \me  b_4\, \bl{\zeta}^3  \ma  b_3\, \bl{\zeta}^4 \,\right)\,,
\end{equation}
where $b_1$, $b_2$, $b_3$ and  $b_4$ are arbitrary complex constants. Besides these eight linearly independent Killing spinors with eigenvalue $\alpha$, the $dS_6$ spacetime also possesses eight Killing spinors with eigenvalue $-\alpha$, the latter being given by
\begin{equation}\label{KS-dS2}
 \widetilde{\bl{\psi}} \eq \Upsilon\,\bl{\psi}\eq \left(\, b_1\, \bl{\chi}_1  \ma b_2\, \bl{\chi}_2 \ma b_3\, h\,\bl{\chi}_3 \ma b_4\, h\,\bl{\chi}_4 \,\,,\,   b_2\,  h\, \bl{\zeta}^1 \me  b_1\,  h\, \bl{\zeta}^2  \ma  b_4\, \bl{\zeta}^3  \me  b_3\, \bl{\zeta}^4 \,\right) \,.
\end{equation}
The scalars constructed from the spinor (\ref{KS-dS}) are given by
$$ \psi^A\,\psi_A \eq 0 \quad, \quad
 \overline{\psi}^A\,\psi_A \eq  \psi^A\,\overline{\psi}_A \eq  h \lef |b_1|^2 \ma |b_2|^2 \ma |b_3|^2 \ma |b_4|^2 \rig \,. $$
Therefore, since $\overline{\psi}^A \psi_A\neq 0$, no Killing spinors of the $dS_6$ spacetime are of the first type described in (\ref{StandardForms}). Indeed, we have proved in Section \ref{Subsection-KStype1} that spinors of the latter type cannot occur in a spacetime whose curvature scalar is positive.

\section{Conclusions}\label{Sec.Conclusion}

After reviewing the spinorial calculus and its index notation in six dimensions, we have obtained how the equations satisfied by Killing-Yano (KY) tensors and closed conformal Killing-Yano (CCKY) tensors are represented in the latter formalism. Moreover, it has been shown how the Killing spinor equation is transcribed to the index approach. These results are summarized in Table \ref{TableEquations}. Then, we have taken advantage of the index formalism to construct all possible symmetry tensors that are descendant from the existence of a Killing spinor, as summed up in Table \ref{TableSymmetryTensors}.  In addition, in Sec. \ref{SecIntegrab.Cond.} we have expressed the integrability conditions necessary for the existence of a Killing spinor in terms of the spinorial representations of the Weyl tensor, the traceless part of the Ricci tensor and the Ricci scalar. Although, at first, it may seem that the index notation is  clumsier than the usual abstract approach, it actually has several advantages. Indeed, once one gets used, the index notation illuminates and facilitates the spinorial manipulations. For instance, the cumbersome Fierz identities are unnecessary in such approach, since these identities are naturally built in the index notation.

The power and the usefulness of the six-dimensional index formalism has been made more explicit in Sec. \ref{Sec.Integration}, in which the Killing spinor equation in the Lorentzian signature has been integrated in some cases. Indeed, we have taken advantage of the index approach to classify the possible algebraic forms that a Killing spinor can have in two categories, as shown in Eq. (\ref{StandardForms2}). This humble classification allowed us to start the integration process from a simple \textit{ansatz} without loosing generality. Particularly, we have proved that in order for a spacetime to admit a Killing spinor its Weyl tensor must have one of the following three algebraically special types according to the CMPP classification: (1) type $O$, namely the Weyl tensor vanishes; (2) type $N$; (3) type $D$. Moreover, in the case of a type $N$ Weyl tensor, the scalar curvature must be negative, while in the case of type $D$ the spacetime must be an Einstein manifold. Further, in Eq. (\ref{Metric1}), we have presented a type $N$ class of spacetimes, with four arbitrary functions in the metric, that possesses a Killing spinor and, therefore, admits a whole tower of hidden symmetries represented by KY and CCKY tensors. Even imposing the Einstein condition for the latter class, there remain three arbitrary functions in the metric, see Eq. (\ref{SolutionEinstein1}). Finally, the Killing spinor equation has been fully integrated in the $dS_6$ and $AdS_6$ spacetimes. Each of these maximally symmetric spacetimes admits sixteen Killing spinors, half of them having eigenvalue $\alpha$ while the other half has eigenvalue $-\alpha$, where $\alpha=\sqrt{\frac{-\,R}{4n(n-1)}}$. The explicit forms of these Killing spinors are given in Eqs. (\ref{KS-AdS}), (\ref{KS-AdS2}), (\ref{KS-dS}), and (\ref{KS-dS2}).

The main physical applications of this work are related to supergravity and string theories, inasmuch as the existence of a Killing spinor in a solution is necessary for the preservation of part of the supersymmetry. The existence of unbroken supersymmetries, in turn, guarantees the stability of the ground state on a spontaneous compactification, although it is not a necessary condition \cite{DuffPope-sugra}. We have investigated the existence of Killing spinors specially in Lorentzian six-dimensional manifolds, which is of relevance for supergravity models and effective theories in six dimensions. A natural continuation of this work that we intend to pursuit in the near future is to analyse theories in the presence of matter fields, in which case the connection adopted in the Killing spinor equation differ from the spinorial extension of the Levi-Civita connection by terms depending on gauge potentials and their field strength. Another path to be taken in a future work is to study compact six-dimensional manifolds of Euclidean signature possessing Killing spinors, which are suitable to be used as internal spaces on compactifications of ten-dimensional fundamental theories.



\section*{Acknowledgments}
I would like to acknowledge the valuable discussions with Bruno Carneiro da Cunha and thank Juliana Magalh\~{a}es Franca for the unswerving support.

\appendix

\section{ Lorentz Transformations in the Spinorial Formalism }\label{AppendixLorentz}

The aim of this appendix is to display all transformations of the spinorial basis $\{\bl{\chi}_1, \bl{\chi}_2, \bl{\chi}_3, \bl{\chi}_4\}$ that preserve the metric. The key to accomplish this is to note that, apart from the multiplicative factor of $1/2$, the metric of the space is represented by $\varepsilon_{ABCD}$, which, raising the indices, can be written as $\varepsilon^{ABCD}\eq \chi_1^{\,[A} \chi_2^{\,B} \chi_3^{\,C} \chi_4^{\,D]}$. Therefore, preserving the metric is tantamount to preserving the skew-symmetric combination $\chi_1^{\,[A} \chi_2^{\,B} \chi_3^{\,C} \chi_4^{\,D]}$. The most general transformation of the basis that does this job is a composition of the  following five transformations.
\begin{equation}\label{NullRotarion1}
 \bl{\chi}_1 \mapsto \bl{\chi}_1 \;,\quad
 \bl{\chi}_2 \mapsto \bl{\chi}_2 + a_1\,\bl{\chi}_1 \quad,\quad
 \bl{\chi}_3 \mapsto \bl{\chi}_3 + a_2\,\bl{\chi}_1  \quad,\quad
 \bl{\chi}_4 \mapsto \bl{\chi}_4 + a_3\,\bl{\chi}_1 \,.
\end{equation}
\begin{equation}\label{NullRotarion2}
 \bl{\chi}_1 \mapsto \bl{\chi}_1 + b_1\,\bl{\chi}_2 \quad,\quad
 \bl{\chi}_2 \mapsto \bl{\chi}_2  \quad,\quad
 \bl{\chi}_3 \mapsto \bl{\chi}_3 + b_2\,\bl{\chi}_2  \quad,\quad
 \bl{\chi}_4 \mapsto \bl{\chi}_4 + b_3\,\bl{\chi}_2 \,.
\end{equation}
\begin{equation}\label{NullRotarion3}
 \bl{\chi}_1 \mapsto \bl{\chi}_1 + c_1\,\bl{\chi}_3 \quad,\quad
 \bl{\chi}_2 \mapsto \bl{\chi}_2 + c_2\,\bl{\chi}_3  \quad,\quad
 \bl{\chi}_3 \mapsto \bl{\chi}_3  \quad,\quad
 \bl{\chi}_4 \mapsto \bl{\chi}_4 + c_3\,\bl{\chi}_3 \,.
\end{equation}
\begin{equation}\label{NullRotarion4}
 \bl{\chi}_1 \mapsto \bl{\chi}_1 + d_1\,\bl{\chi}_4 \quad,\quad
 \bl{\chi}_2 \mapsto \bl{\chi}_2 + d_2\,\bl{\chi}_4  \quad,\quad
 \bl{\chi}_3 \mapsto \bl{\chi}_3 + d_3\,\bl{\chi}_4  \quad,\quad
 \bl{\chi}_4 \mapsto \bl{\chi}_4  \,.
\end{equation}
\begin{equation}\label{Boost}
 \bl{\chi}_1 \mapsto z_1 z_2 z_3 \,\bl{\chi}_1 \quad,\quad
  \bl{\chi}_2 \mapsto \frac{z_1}{z_2 z_3}  \,\bl{\chi}_2 \quad,\quad
  \bl{\chi}_3 \mapsto \frac{z_2}{z_1 z_3}  \,\bl{\chi}_3 \quad,\quad
 \bl{\chi}_4 \mapsto \frac{z_3}{z_1 z_2}  \,\bl{\chi}_4  \,.
\end{equation}
Where the parameters $z_i$, $a_i$, $b_i$, $c_i$, $d_i$ are 15 arbitrary complex parameters. The transformation (\ref{NullRotarion1}) preserves the spinor $\bl{\chi}_1$. This spinor is the pure spinor associated to the maximally isotropic distribution generated by the null vectors
 $\{\bl{e}_1, \bl{e}_2, \bl{e}_3\}$. Therefore, these vector fields are preserved by the transformation (\ref{NullRotarion1}). Such transformation is the six-dimensional analogue of the so-called null rotation in four dimensions. Likewise, the transformations (\ref{NullRotarion2}), (\ref{NullRotarion3}) and (\ref{NullRotarion4}) preserve the maximally isotropic distributions associated to the pure spinors $\bl{\chi}_2$, $\bl{\chi}_3$ and $\bl{\chi}_4$, respectively, and should also be understood as null rotations. Finally, the transformation (\ref{Boost}) implements a scaling on the vector fields of the null frame, this is the analogue of the boost transformation in four dimensions.

However, not all these transformations preserve the conjugation relations (\ref{Conjugation}) that assure the Lorentzian signature. Indeed, in order for such conjugation relations to remain valid, these parameters must satisfy some reality conditions. For instance, if
$$  z_1 = z_1^* \;\;,\;\;  z_2 \,z_2^* = 1  \;\;,\;\; z_3 \,z_3^* = 1 \;\;,  $$
while the other parameters are zero, we have an orthogonal transformation that preserves the Lorentzian signature. A convenient way to write the transformations that preserve both the inner products and the Lorentz signature is a composition of the following three transformations
\begin{equation}\label{LorentzTransf1}
 \left\{
   \begin{array}{ll}
    \bl{\chi}_1 \mapsto  \lambda \lef e^{i\,\phi_1} \cos\theta_1\, \bl{\chi}_1  \ma e^{i\,\phi_2} \sin\theta_1\, \bl{\chi}_2 \rig   \quad,&\quad
\bl{\chi}_2 \mapsto \lambda \lef -\,e^{-i\,\phi_2} \sin\theta_1\, \bl{\chi}_1  \ma e^{-i\,\phi_1} \cos\theta_1\, \bl{\chi}_2 \rig   \;, \\
    \bl{\chi}_3 \mapsto  \frac{1}{\lambda} \lef e^{i\,\phi_3} \cos\theta_2\,\, \bl{\chi}_3  \me e^{i\,\phi_4} \sin\theta_2\,\, \bl{\chi}_4 \rig   \quad, & \quad
\bl{\chi}_4 \mapsto \frac{1}{\lambda} \lef e^{-i\,\phi_4} \sin\theta_2\,\, \bl{\chi}_3  \ma e^{-i\,\phi_3} \cos\theta_2\,\, \bl{\chi}_4 \rig   \,.
   \end{array}
 \right.
 \end{equation}
\begin{equation}\label{LorentzTransf2}
 \bl{\chi}_1 \mapsto \bl{\chi}_1 \ma a\, \bl{\chi}_3  \ma   b\, \bl{\chi}_4   \quad,\quad
\bl{\chi}_2 \mapsto \bl{\chi}_2 \ma b^*\, \bl{\chi}_3  \me   a^*\, \bl{\chi}_4   \quad,\quad
\bl{\chi}_3 \mapsto  \bl{\chi}_3  \quad,\quad  \bl{\chi}_4 \mapsto  \bl{\chi}_4 \,.
\end{equation}
\begin{equation}\label{LorentzTransf3}
  \bl{\chi}_1 \mapsto  \bl{\chi}_1  \quad,\quad  \bl{\chi}_2 \mapsto  \bl{\chi}_2   \quad,\quad
\bl{\chi}_3 \mapsto \bl{\chi}_3 \me c\, \bl{\chi}_1  \ma   d\, \bl{\chi}_2   \quad,\quad
\bl{\chi}_4 \mapsto \bl{\chi}_4 \ma d^*\, \bl{\chi}_1  \ma  c^*\, \bl{\chi}_2   \,.
\end{equation}
Where the $\phi$'s, $\theta$'s and $\lambda$  are real parameters, while $a$, $b$, $c$, $d$ are complex, forming a total of 15 real variables, as it should be, since the Lorentz group $SO(1,5)$ has 15 generators. These transformations on the basis of positive chirality spinors  induce a modification on the dual basis of negative chirality spinors. The latter transformations can be obtained from the following relations:
\begin{align}
  \zeta^1_{\,A} \eq \varepsilon_{ABCD}\,\chi_2^{\,B} \chi_3^{\,C}  \chi_4^{\,D} \quad , & \quad
\zeta^2_{\,B} \eq \varepsilon_{ABCD}\,\chi_1^{\,A} \chi_3^{\,C}  \chi_4^{\,D} \;,  \label{NegativeBasis} \\
  \zeta^3_{\,C} \eq \varepsilon_{ABCD}\,\chi_1^{\,A} \chi_2^{\,B}  \chi_4^{\,D}  \quad , & \quad
\zeta^4_{\,D} \eq \varepsilon_{ABCD}\,\chi_1^{\,A} \chi_2^{\,B}  \chi_3^{\,C}  \,.\nonumber
\end{align}
Indeed, inserting Eqs. (\ref{LorentzTransf1}), (\ref{LorentzTransf2}) and (\ref{LorentzTransf3}) into Eq. (\ref{NegativeBasis}), we eventually arrive at the following transformation rules respectively:
\begin{equation}\label{LorentzTransf1neg}
 \left\{
   \begin{array}{ll}
    \bl{\zeta}_1 \mapsto  \frac{1}{\lambda} \lef e^{-i\,\phi_1} \cos\theta_1\, \bl{\zeta}_1  \ma e^{-i\,\phi_2} \sin\theta_1\, \bl{\zeta}_2 \rig   \quad,&\quad
\bl{\zeta}_2 \mapsto \frac{1}{\lambda} \lef -\,e^{i\,\phi_2} \sin\theta_1\, \bl{\zeta}_1  \ma e^{i\,\phi_1} \cos\theta_1\, \bl{\zeta}_2 \rig   \;, \\
    \bl{\zeta}_3 \mapsto  \lambda \lef e^{-i\,\phi_3} \cos\theta_2\,\, \bl{\zeta}_3  \me e^{-i\,\phi_4} \sin\theta_2\,\, \bl{\zeta}_4 \rig   \quad, & \quad
\bl{\zeta}_4 \mapsto \lambda \lef e^{i\,\phi_4} \sin\theta_2\,\, \bl{\zeta}_3  \ma e^{i\,\phi_3} \cos\theta_2\,\, \bl{\zeta}_4 \rig   \,.
   \end{array}
 \right.
 \end{equation}
\begin{equation}\label{LorentzTransf3neg}
  \bl{\zeta}_1 \mapsto  \bl{\zeta}_1  \quad,\quad  \bl{\zeta}_2 \mapsto  \bl{\zeta}_2   \quad,\quad
\bl{\zeta}_3 \mapsto \bl{\zeta}_3 \me a\, \bl{\zeta}_1  \me   b^*\, \bl{\zeta}_2   \quad,\quad
\bl{\zeta}_4 \mapsto \bl{\zeta}_4 \me b\, \bl{\zeta}_1  \ma  a^*\, \bl{\zeta}_2   \,.
\end{equation}
\begin{equation}\label{LorentzTransf2neg}
 \bl{\zeta}_1 \mapsto \bl{\zeta}_1 \ma c\, \bl{\zeta}_3  \me   d^*\, \bl{\zeta}_4   \quad,\quad
\bl{\zeta}_2 \mapsto \bl{\zeta}_2 \me d\, \bl{\zeta}_3  \me   c^*\, \bl{\zeta}_4   \quad,\quad
\bl{\zeta}_3 \mapsto  \bl{\zeta}_3  \quad,\quad  \bl{\zeta}_4 \mapsto  \bl{\zeta}_4 \,.
\end{equation}

These Lorentz transformations can be quite useful for simplifying the algebraic form of spinors. For instance, let $\bl{\psi} = (\psi^A, \psi_A)$ be a general spinor with $\psi^A \neq 0$. Then, without loss of generality, one can always align our spinorial basis in such a way that  $\psi^A = \chi_1^{\,A}$. Thus, a spinor that is non-chiral can always be written as
$$ \bl{\psi} \eq \lef \bl{\chi}_1 \,,\, f_1 \,\bl{\zeta}^1 +  f_2 \,\bl{\zeta}^2 + f_3 \,\bl{\zeta}^3 + f_4 \,\bl{\zeta}^4  \rig \,.$$
Then, if either $f_1$ or $f_2$ is different from zero, namely $\psi^A \psi_A \neq 0$ or  $\psi^A \overline{\psi}_A \neq 0$, one can get rid of the coefficients $f_3$ and $f_4$ by a proper choice of frame. Indeed, performing the transformations (\ref{LorentzTransf3}) and (\ref{LorentzTransf2neg}) with
$$  c \eq \frac{f_2\,f_4^* \me f_1^*\,f_3}{|f_1|^2 \ma |f_2|^2} \quad \textrm{ and } \quad  d \eq \frac{f_2^*\,f_3 \ma f_1\,f_4^*}{|f_1|^2 \ma |f_2|^2}  \,, $$
it follows that the spinor $\bl{\psi}$ is written in the new frame as
\begin{equation}\label{AlgebraicSimp1}
 \bl{\psi} \eq \lef \bl{\chi}_1 \,,\, f_1 \,\bl{\zeta}^1 +  f_2 \,\bl{\zeta}^2  \rig \,.
\end{equation}
Therefore, we have proved that given an arbitrary non-chiral spinor $\bl{\psi}$ then if either $\psi^A \psi_A \neq 0$ or  $\psi^A \overline{\psi}_A \neq 0$ then one can always align the spinorial frame in such a way that $\bl{\psi}$ assumes the algebraic form presented in Eq. (\ref{AlgebraicSimp1}). This represents a great simplification in the algebraic form of the spinor, which, otherwise, would have eight coefficients instead of just two. Analogously, if $\bl{\psi}$ is a non-chiral spinor such that both $\psi^A \psi_A$ and  $\psi^A \overline{\psi}_A$ vanish then it is always possible to conveniently align the spinor in such a way that
\begin{equation}\label{AlgebraicSimp2}
  \bl{\psi} \eq \lef \bl{\chi}_1 \,,\, f \,\bl{\zeta}^4   \rig \,.
\end{equation}
Moreover, one can always absorb the complex phase in $f$ by properly choosing the parameter $\phi_3$ in (\ref{LorentzTransf1neg}). Thus, we can assume that $f$ is a real function.

\section{ The Spinorial Connection }\label{AppendixConnection}

Since the chirality matrix $\Upsilon$ is covariantly constant, it follows that the covariant derivative preserves the chirality of a spinor. Therefore the covariant derivative of a spinor of positive chirality $\psi^A$ is another spinor of positive chirality, so that the covariant derivative keeps the spinorial index up:
\begin{equation}\label{CovD1}
   \nabla_{AB} \, \psi^C \eq \partial_{AB}\, \psi^C \ma (\Omega_{AB})^C_{\ph{C}D}\,\psi^D \,.
\end{equation}
Analogously, the down index of a spinor of negative chirality is also kept down by the action of the covariant derivative. More precisely, we have that
\begin{equation}\label{CovD2}
 \nabla_{AB} \, \psi_C \eq \partial_{AB}\, \psi_C \me (\Omega_{AB})^D_{\ph{D}C}\,\psi_D \,.
\end{equation}
The latter expression can be extracted from (\ref{CovD1}) by using the Leibniz rule along with the fact that $\psi^A\psi_A$ is a scalar, so that
$$ \nabla_{AB} \,(\psi^C \,\psi_C) \eq \partial_{AB} \,(\psi^C \,\psi_C)\,.  $$
The relation between the spinorial connection $(\Omega_{AB})^C_{\ph{C}D}$ and the connection of the tangent bundle $\omega_{ab}^{\ph{ab}c}$ can be found by means of Eq. (\ref{CovDSpinors}), once a representation for the matrices $\gamma_a$ has been introduced. For this purpose, let us present an specific representation of the null frame $\{\bl{e}_a\}$, i.e. a frame whose mutual inner product relations are the ones presented in Eq. (\ref{NullFrameInnerP}). Choosing the spinorial basis to be  $\{\bl{\chi}_1,\,\bl{\chi}_2,\,\bl{\chi}_3,\,\bl{\chi}_4,\,\bl{\zeta}^1,\,\bl{\zeta}^2,\,\bl{\zeta}^4,\,\bl{\zeta}^4\}$,
it follows from the definitions (\ref{NullFrame1}) and  (\ref{NullFrame2}) and from the identity (\ref{CliffordAction6D})  that the null frame have the following matrix representation.
\small
$$  \gamma_1 \eq \left[
\begin{array}{cccccccc}
 0 & 0 & 0 & 0 & 0 & 1 & 0 & 0 \\
 0 & 0 & 0 & 0 & -1 & 0 & 0 & 0 \\
 0 & 0 & 0 & 0 & 0 & 0 & 0 & 0 \\
 0 & 0 & 0 & 0 & 0 & 0 & 0 & 0 \\
 0 & 0 & 0 & 0 & 0 & 0 & 0 & 0 \\
 0 & 0 & 0 & 0 & 0 & 0 & 0 & 0 \\
 0 & 0 & 0 & -1 & 0 & 0 & 0 & 0 \\
 0 & 0 & 1 & 0 & 0 & 0 & 0 & 0 \\
\end{array}
\right] \;,\;
\gamma_4 \eq \left[
\begin{array}{cccccccc}
 0 & 0 & 0 & 0 & 0 & 0 & 0 & 0 \\
 0 & 0 & 0 & 0 & 0 & 0 & 0 & 0 \\
 0 & 0 & 0 & 0 & 0 & 0 & 0 & 1 \\
 0 & 0 & 0 & 0 & 0 & 0 & -1 & 0 \\
 0 & -1 & 0 & 0 & 0 & 0 & 0 & 0 \\
 1 & 0 & 0 & 0 & 0 & 0 & 0 & 0 \\
 0 & 0 & 0 & 0 & 0 & 0 & 0 & 0 \\
 0 & 0 & 0 & 0 & 0 & 0 & 0 & 0 \\
\end{array}
\right] \;, $$
$$
\gamma_2 \eq  \left[
\begin{array}{cccccccc}
 0 & 0 & 0 & 0 & 0 & 0 & 1 & 0 \\
 0 & 0 & 0 & 0 & 0 & 0 & 0 & 0 \\
 0 & 0 & 0 & 0 & -1 & 0 & 0 & 0 \\
 0 & 0 & 0 & 0 & 0 & 0 & 0 & 0 \\
 0 & 0 & 0 & 0 & 0 & 0 & 0 & 0 \\
 0 & 0 & 0 & 1 & 0 & 0 & 0 & 0 \\
 0 & 0 & 0 & 0 & 0 & 0 & 0 & 0 \\
 0 & -1 & 0 & 0 & 0 & 0 & 0 & 0 \\
\end{array}
\right] \;,\;
\gamma_5 \eq  \left[
\begin{array}{cccccccc}
 0 & 0 & 0 & 0 & 0 & 0 & 0 & 0 \\
 0 & 0 & 0 & 0 & 0 & 0 & 0 & -1 \\
 0 & 0 & 0 & 0 & 0 & 0 & 0 & 0 \\
 0 & 0 & 0 & 0 & 0 & 1 & 0 & 0 \\
 0 & 0 & -1 & 0 & 0 & 0 & 0 & 0 \\
 0 & 0 & 0 & 0 & 0 & 0 & 0 & 0 \\
 1 & 0 & 0 & 0 & 0 & 0 & 0 & 0 \\
 0 & 0 & 0 & 0 & 0 & 0 & 0 & 0 \\
\end{array}
\right] \;, $$
$$ \gamma_3 \eq \left[
\begin{array}{cccccccc}
 0 & 0 & 0 & 0 & 0 & 0 & 0 & 1 \\
 0 & 0 & 0 & 0 & 0 & 0 & 0 & 0 \\
 0 & 0 & 0 & 0 & 0 & 0 & 0 & 0 \\
 0 & 0 & 0 & 0 & -1 & 0 & 0 & 0 \\
 0 & 0 & 0 & 0 & 0 & 0 & 0 & 0 \\
 0 & 0 & -1 & 0 & 0 & 0 & 0 & 0 \\
 0 & 1 & 0 & 0 & 0 & 0 & 0 & 0 \\
 0 & 0 & 0 & 0 & 0 & 0 & 0 & 0 \\
\end{array}
\right] \;,\;
\gamma_6 \eq \left[
\begin{array}{cccccccc}
 0 & 0 & 0 & 0 & 0 & 0 & 0 & 0 \\
 0 & 0 & 0 & 0 & 0 & 0 & 1 & 0 \\
 0 & 0 & 0 & 0 & 0 & -1 & 0 & 0 \\
 0 & 0 & 0 & 0 & 0 & 0 & 0 & 0 \\
 0 & 0 & 0 & -1 & 0 & 0 & 0 & 0 \\
 0 & 0 & 0 & 0 & 0 & 0 & 0 & 0 \\
 0 & 0 & 0 & 0 & 0 & 0 & 0 & 0 \\
 1 & 0 & 0 & 0 & 0 & 0 & 0 & 0 \\
\end{array}
\right]\;.$$
\normalsize
Inserting this representation into Eq. (\ref{CovDSpinors}), that relates the spinorial connection with the tangent bundle connection, we eventually arrive at the following expression for the object $(\Omega_{AB})^C_{\ph{C}D}$:
\small
\begin{equation}\label{Connection6D}
  (\Omega_a)^C_{\ph{C}D} = \left[
      \begin{array}{cccc}
\frac{-1}{2}(\omega_{a4}^{\ph{a4}4} + \omega_{a5}^{\ph{a5}5}+\omega_{a6}^{\ph{a6}6}) &  -\omega_{a5}^{\ph{a5}3}
       & \omega_{a4}^{\ph{a4}3} &-\omega_{a4}^{\ph{a4}2} \\
        \omega_{a2}^{\ph{a2}6} & \frac{-1}{2}(\omega_{a4}^{\ph{a4}4} -\omega_{a5}^{\ph{a5}5} - \omega_{a6}^{\ph{a6}6})
          & -\omega_{a4}^{\ph{a4}5} &  -\omega_{a4}^{\ph{a4}6}  \\
        -\omega_{a1}^{\ph{a1}6} & \omega_{a1}^{\ph{a1}2}
          & \frac{1}{2}(\omega_{a4}^{\ph{a4}4} - \omega_{a5}^{\ph{a5}5}+\omega_{a6}^{\ph{a6}6}) & -\omega_{a5}^{\ph{a5}6} \\
        \omega_{a1}^{\ph{a1}5} & \omega_{a1}^{\ph{a1}3}
       & \omega_{a2}^{\ph{a2}3} & \frac{1}{2}(\omega_{a4}^{\ph{a4}4} + \omega_{a5}^{\ph{a5}5}-\omega_{a6}^{\ph{a6}6}) \\
      \end{array}
    \right]
\end{equation}
\normalsize
Where it is worth recalling that $\omega_{ab}^{\ph{ab}c}$ is defined by the relation $\nabla_a \bl{e}_b = \omega_{ab}^{\ph{ab}c}\bl{e}_c$ and that $\omega_{abc} \eq -\omega_{acb}$. The conversion between the vectorial index $a$ in $(\Omega_a)^C_{\ph{C}D}$ and the spinorial indices $AB$ in $(\Omega_{AB})^C_{\ph{C}D}$ is easily done by means of Eq. (\ref{NullFrame1}). Thus, for example, Eq. (\ref{Connection6D}) yields
$$ \zeta^4_{\;A}\lef \nabla_{23}\,\chi_1^{\;A} \rig \,\equiv \,  \zeta^4_{\;A}\lef \nabla_{6}\,\chi_1^{\;A} \rig \eq (\Omega_6)^4_{\ph{4}1} \eq
\omega_{61}^{\ph{61}5} \eq \omega_{6\ph{4}2}^{\ph{6}4}   \eq - \,\omega_{62}^{\ph{62}4} \,. $$


\end{document}